\documentclass[10pt,a4paper,twocolumn,english]{IEEEtran}
\usepackage[T1]{fontenc}
\usepackage{fancyhdr}
\usepackage{babel}
\usepackage{float}
\usepackage{url}
\usepackage{amsmath}
\usepackage{amssymb}
\usepackage{graphicx}
\usepackage[unicode=true,
 bookmarks=true,bookmarksnumbered=true,bookmarksopen=true,bookmarksopenlevel=1,
 breaklinks=false,pdfborder={0 0 0},backref=false,colorlinks=false]
 {hyperref}
\hypersetup{pdftitle={Your Title},
 pdfauthor={Your Name},
 pdfpagelayout=OneColumn, pdfnewwindow=true, pdfstartview=XYZ, plainpages=false}
\usepackage{breakurl}

\makeatletter

\providecommand{\tabularnewline}{\\}
\floatstyle{ruled}
\newfloat{algorithm}{tbp}{loa}
\providecommand{\algorithmname}{Algorithm}
\floatname{algorithm}{\protect\algorithmname}

\ifCLASSOPTIONcompsoc
\usepackage[caption=false,font=normalsize,labelfont=sf,textfont=sf]{subfig}
\else
\usepackage[caption=false,font=footnotesize]{subfig}
\fi
\makeatletter\let\ps@plain\ps@fancy\makeatother

\makeatother

\begin{document}

\title{Interference-Cooperation in Multi-User/Multi-Operator Receivers}

\author{Syed Hassan Raza Naqvi$^{(*)}$, Umberto Spagnolini\\
Dipartimento di Elettronica, Informazione e Bioingegneria (DEIB)\\
Politecnico di Milano\\
P.zza L. da Vinci 32, I - 20133 Milano, Italy. \\
Email: {\{Syedhassan.Naqvi,Umberto.Spagnolini\}@polimi.it} \\
 $^{(*)}$ {\small{}Corresponding author}}

\maketitle
\thispagestyle{fancy}
\begin{abstract}
In a multi-user scenario where users belong to different operators,
any interference mitigation method needs unavoidably some degree of
cooperation among service providers. In this paper we propose a cooperation
strategy based on the exchange of mutual interference among operators,
rather than of decoded data, to let every operator to recover an augmented
degree of diversity either for channel estimation and multi-user detection.
In xDSL scenario where multiple operators share the same cable binder
the interference-cooperation (IC) approach outperforms data-exchange
methods and preserves to certain degree the privacy of the users as
signals can be tailored to prevent each operator to infer parameters
(channel and data) of the users from the other operators.

The IC method is based on Expectation Maximization estimation shaped
to account for the degree of information that each operator can exchange
with the others during the two steps of multi-user channel estimation
and multi-user detection. Convergence of IC is guaranteed into few
iterations and it does not depend on the structure of the interference.
IC performance attains those of centralized receivers (i.e., one fusion-center
that collects all the received signals from all the users/operators),
with some loss when in heavily interfered multi-user channel such
as in twisted-pair communications allocated beyond 50-100MHz spectrum. \end{abstract}


\section{Introduction}

\IEEEPARstart{I}{nterference} mitigation is a largely investigated
topic in communication systems for wired and wireless technology.
The increasing demand of data-intensive services in fixed and mobile
communication brings new challenges for interference cancellation,
and service providers (SPs) are looking for solutions that not only
meet current, but mostly future requirements, still preserving the
downward compatibility.

\begin{figure}[h]
\centering \includegraphics[width=0.5\textwidth]{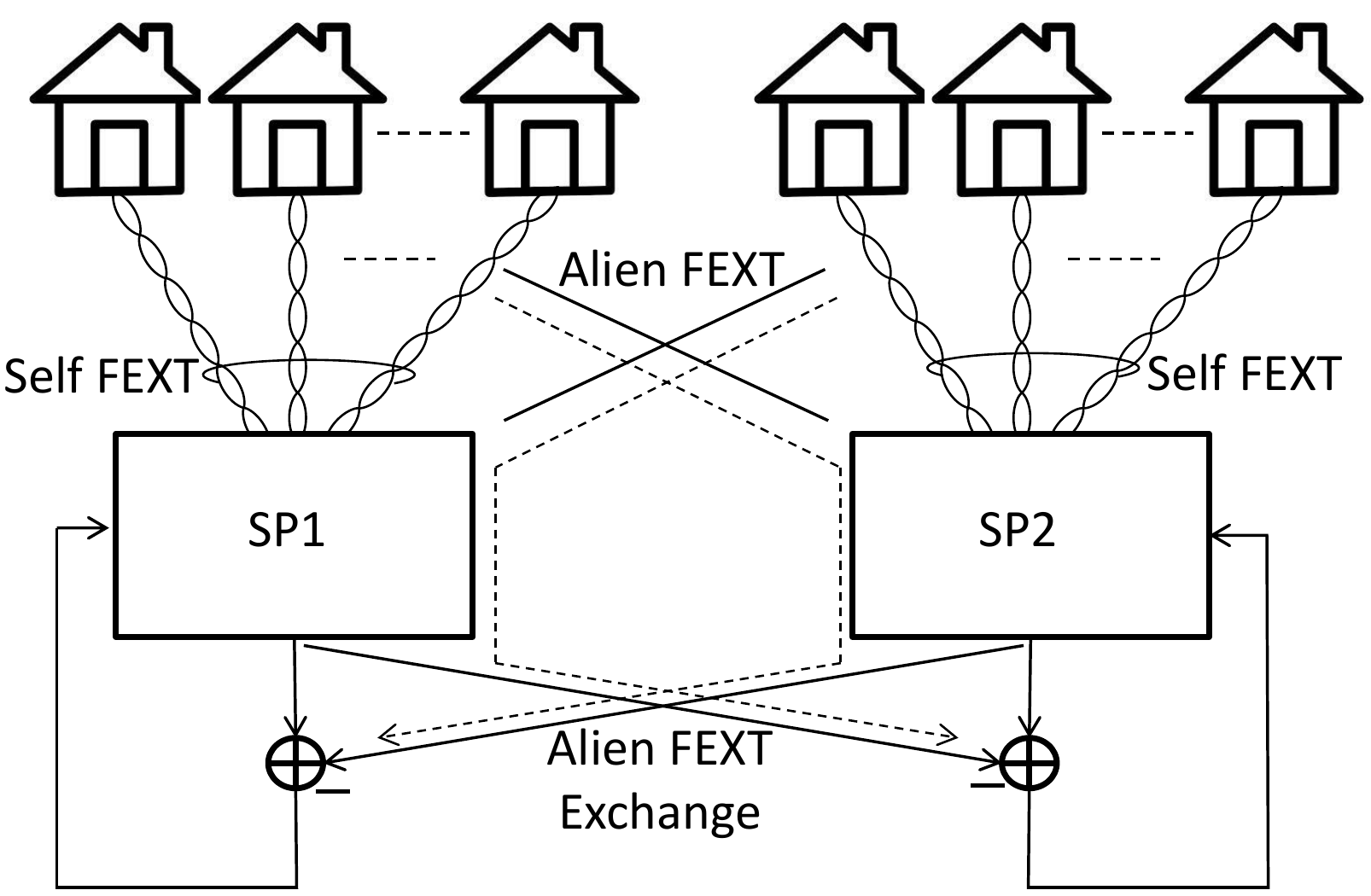}
\caption{Interference Cooperation (IC) for multi-operator xDSL system.}
\label{fig:Fig.1}
\end{figure}

Existing copper-wire infrastructure to provide net-bidirectional data
rate of up to 200Mbps can use short loop lengths and high frequencies
as in ITU-T G.993.2 recommendation \cite{cit1-1}. Single cable binder
can contain up to hundreds twisted pairs and these can be shared by
multiple coexisting SPs. In multi-pair cable, crosstalk is the dominant
impairment caused by the capacitive and inductive coupling among the
twisted pairs. Techniques have been proposed to mitigate near-end
crosstalk (NEXT) such as spectral shaping and frequency division duplexing
\cite{cit2}, \cite{cit3}. However, in digital subscriber line (DSL)
the data rate is still limited by the far-end crosstalk (FEXT) that
can be mitigated only by appropriate interference cancellation methods.
More specifically, the performance is improved by using multi-user
processing for FEXT cancellation, or vectoring \cite{cit4}. DSL Access
Multiplexer (DSLAM) hosts signal processing units to mitigate upstream
FEXT generated within a vectored group belonging to the same SP. In
common DSL scenario, non-vectored lines coexist along with the vectored
groups in a same cable binder, and this prevents reliable FEXT control.
Referring to multi-operator scenario in Fig.\ref{fig:Fig.1}, vectoring
technique is very efficient to cancel self-FEXT (i.e., FEXT of the
same SP), but it has no control of FEXT due to the non-vectored lines
and from other vectored group in the same cable binder (so called
alien-FEXT) when twisted-pairs are shared among non-cooperating SPs.
This problem is even more challenging in next generation G.Fast standard
where the bandwidth is even larger than 100MHz \cite{cit5} over 50-200m
cable length and alien-FEXT induced from just one temporary line-mismatch
could loose all the benefits of vectoring at the expenses of energy
\cite{ICCC2014}.

In this scenario, centralized vectoring where FEXT mitigation is controlled
by one single processing unit for multiple SPs would be beneficial
but it is unfeasible due to the regulations of physical unbundling
\cite{cit6}. In addition, due to the privacy issue, SPs are not prone
to exchange each other data to ease interference cancellation, even
if in turn from this exchange there would be a benefit for all. Dynamic
spectrum management methods could provide the same benefit as vectored
transmission in the mix vectored and non-vectored (or equivalently
non-cooperating SPs) cable environment \cite{cit7}. Noise decorrelation
technique can handle alien-FEXT as additive structured noise, at the
price of additional initialization (to estimate noise correlation)
and adaptation (for time varying alien-FEXT scenario) algorithms \cite{cit8}.
If in vectored groups the non-vectored lines are unmanaged, the benefits
of vectoring degrade rapidly \cite{cit9}. The way to handle compatibility
between vectored and non-vectored group lines is still an open issue
\cite{cit8}.

Focus of this paper is the co-operation among SPs as envisioned in
\cite{zidane2013vectored} to take the full advantage of vectoring
in multi-operator settings. Cooperation among multiple SPs (Fig.\ref{fig:Fig.1})
being one alien-FEXT of the other is proposed here by avoiding the
exchange of any sensitive data among SPs to guarantee that none of
the other SPs can recover the information (either channel matrix and
encoded data stream) of own users. The interference cooperation (IC)
approach is based on the exchange of alien-FEXT only that is purposely
removed as sketched in Fig.\ref{fig:Fig.1}. The iterative ICs that
account either for multi-user channel estimation and multi-user detection
are based on Expectation and Maximization (EM) estimation methods
adapted to FEXT mitigation from cellular communications \cite{cit10}\cite{cit11}\cite{cit12}.
In this paper, these iterative EM methods are reformulated by guaranteeing
that every SP exchanges the interference with others after stripping
the information depending on the own users to have an IC that preserves
the sensible information. Given the linearity of the communication
model, multiple-channel estimation and multi-user detection are handled
using the same principle, with slightly differences due to the peculiarities
of the two processing steps. Even if the signaling of the inter-operator
interference seems to be a trivial extension of EM methods, the specificities
of interference cooperation need to address the overall multi-user
channel estimation (comprehensive of alien-FEXT) and detection as
a new problem.

Benefits from cooperation in multiple receiver systems was established
based on information theoretic capacity of multi-receiver cellular
network \cite{cit13,cit14}, showing that the capacity loss due to
inter-cell interference (i.e., alien-FEXT in wired systems) could
be eliminated by cooperating base stations (BSs). In view to densify
the BSs, multiple remote radio equipment are connected to a central
BS through high speed link to get the benefit of central processing
for up-link inter-cell interference mitigation \cite{cit17}. Interference
alignment and cancellation has been proposed and validated in field
trials \cite{cit18} but still based on exchange of encoded data.
Interference cancellation based on iterative subtraction of interfering
signals in cooperative BS clusters is in \cite{cit20,cit21}. Distributed
BS co-operation scheme based on soft-combining for multi-user multi-cell
system is in \cite{cit22}. In all these schemes, the degree of co-operation
is controlled by limiting the backhaul capacity. Mixed soft and hard
information exchange among multiple BSs for interference cancellation
is in \cite{cit23}. All these methods, in addition to \cite{cit25,cit27,cit28},\cite{cit31},
are excellent references to highlight the active research in the field
of interference cancellation when multiple processing units need to
cooperate by exchanging decoded (still sensitive) data to increase
the spectrum efficiency in wireless and wired systems. However, this
paper moves on a different conceptual setting where the exchanging
among SPs is only the interference from received signals after iteratively
stripping own sensitive data, or data mixed with an unknown mixing
matrix acting as randomized multiplicative data perturbation \cite{Privacy Preserving}.

Security and privacy are primary concerns in communication systems.
Providing enough privacy should be also a challenge in any co-operative
interference mitigation technique in multi-operator scenario due to
the exchange of subscriber's private information for the purpose of
canceling the interference. Conventionally, security is provided at
network layer for the well-known security threats e.g. eavesdropping,
man-in-the-middle attack etc. \cite{cit32,cit33}, but the concept
of security is changing in the advanced communication system design.
Examples of attacks on femtocells are in \cite{cit34}. Recently,
research community is giving more attention to physical layer security.
Stochastic geometry approach for physical layer security in cellular
system has been proposed in \cite{cit35}. This paper is motivated
by the need to do all necessary to prevent the other SPs to decode
the information on own users still having a benefit from the mutual
SPs cooperation.

\subsection{Contribution}

Interference cooperation (IC) for multi-operator is the same for cellular
and wired system, but in this paper we specialize the method for xDSL
system, even if extension to multi-cell processing is conceptually
straightforward with minor adaptations (not covered here). To better
highlight the contribution, let us consider Fig.\ref{fig:Fig.1} for
the simple case of two SPs with one user each labeled as 1 and 2 (channel
estimation would be similar except for larger size systems, see Section
III). Signal model is
\begin{eqnarray*}
y_{1} & = & h_{11}x_{1}+h_{21}x_{2}+w_{1}\mathbf{=g}_{1}^{T}\mathbf{x}+w_{1}\\
y_{2} & = & h_{12}x_{1}+h_{22}x_{2}+w_{2}=\mathbf{g}_{2}^{T}\mathbf{x}+w_{2}
\end{eqnarray*}
where SP 1 and 2 receive the interference from the other ($h_{21}x_{2}$
and $h_{12}x_{1}$), and $w_{1},w_{2}$ denote the additive white
Gaussian noise (AWGN). In conventional data-exchange methods each
SP is aware (or estimated separately) of the channel vector ${\bf g}_{1}=\left[h_{11},h_{21}\right]^{T}$
and ${\bf g}_{2}=\left[h_{12},h_{22}\right]^{T}$, symbols $x_{1}$
and $x_{2}$ are locally decoded as $\hat{x}_{1}$ and $\hat{x}_{2}$
(by SP 1 and 2, respectively) and iteratively exchanged to remove
the crosstalk as $y_{1}-h_{21}\hat{x}_{2}$ to estimate $x_{1}$ by
SP1, and similarly for SP2 with $y_{2}-h_{12}\hat{x}_{1}$.

In the proposed IC method, the overall system is modeled as:
\[
\mathbf{y}=\left[\begin{array}{c}
y_{1}\\
y_{2}
\end{array}\right]=\mathbf{h}_{1}x_{1}+\mathbf{h}_{2}x_{2}+\mathbf{w}
\]
with $\mathbf{h}_{1}=[h_{11},h_{12}]^{T}$ and $\mathbf{h}_{2}=[h_{21},h_{22}]^{T}$
the channels known to SP1 and SP2, respectively. The interference
is exchanged at every iteration (i.e., based on the decoded $\hat{x}_{1}$,
the residual $y_{1}-h_{11}\hat{x}_{1}$ and $h_{12}\hat{x}_{1}$are
sent $1\rightarrow2$ that use by SP2 for decoding ${\hat{x}_{2}}$;
a similar reasoning holds true for signaling $2\rightarrow1$) to
yield the $2\times1$ set of observations $\mathbf{y}-\mathbf{h}_{2}\hat{x}_{2}$
and $\mathbf{y}-\mathbf{h}_{1}\hat{x}_{1}$ at SP 1 and 2 to decode
$\hat{x}_{1}$ and $\hat{x}_{2}$ using the locally known channels
$\mathbf{h}_{1}$ and $\mathbf{h}_{2}$. Differently from conventional
iterative receivers where each exchange of data is to guarantee that
the other processing units can reduce the interference, here each
interference exchange favors the other SPs. Even if interference exchange
seems less efficient than encoded data exchange, this paper proves
that the iterative method solves implicitly the centralized problem
of data detection (and similarly for channel estimation) by reorganizing
the interference exchange to carry out Jacobi iterations distributed
over SPs, that in turn converge to the centralized interference cancellation
method within few iterations at price of signaling overhead. The convergence-rate
of IC depends on the degree of interference that is faster when the
channel matrix is diagonal dominant, but still within 3-5iterations.
Even if the exchange among SPs is interference-based to preserve the
privacy on own users, there could be a way for the other SPs to infer
the information on the own users (e.g., SP1 can extract $h_{21}\hat{x}_{2}$
after the stripping of SP2, but still mixed with unknown $h_{21}$
and $\hat{x}_{2}$) by using blind separation methods \cite{Blind-MUD-Wang-Poor,Blind-MUD-Verd=0000F9}.
However, blind-estimation methods need strict assumptions (e.g., known
constellations and/or well algebraic structure of the channel) that
make them unfeasible in practical xDSL systems (see Section III).

Paper is organized as follows. Section II illustrates the system model
for multi-operator scenarios with self-FEXT and alien-FEXT. Section
III describes the co-operative channel estimation using iterative
information shared among the SPs that implements the EM method distributed
among different SPs with partial information on the interference.
Section IV describes the cooperative multi-user detection (MUD) still
using iterative EM with MMSE estimator for multiple-operators. Numerical
validations are in Section V, while Section VI draws some conclusions.

\section{System Model for Multi-operator Receiver}

System model is for \textit{K}-operators scenario with the channel
matrices (or channels in short) for self- and alien-FEXT.

\subsection{Multi-operator System Model}

\begin{figure}[h]
\centering \includegraphics[width=0.5\textwidth]{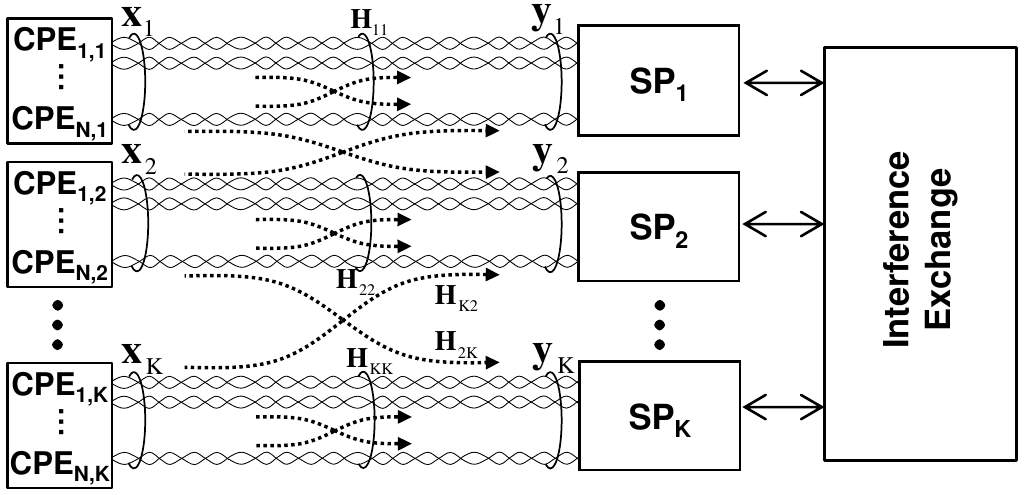}
\caption{System model for cooperative xDSL with $K$ SPs and $N$ CPEs each}
\label{fig:Fig4}
\end{figure}

Multi-operator xDSL system for upstream is shown in Fig.\ref{fig:Fig4},
where multiple service providers (SPs) sharing the same cable binder
have $N$ customer premises equipments (CPEs) labeled as $\text{CP}\text{E}_{1,i},\text{CP}\text{E}_{2,i},\ldots,\text{CP}\text{E}_{N,i}$
that are mutually interfering one another. To have an analytically
tractable problem, here we assume that all the $K$ SPs have the same
number of CPEs mutually synchronized with the same upstream frame
structure without any frequency drift one another so that any inter-carrier
interference can be neglected, and we can employ a system model that
accounts for FEXT on each carrier independently.

Fig.\ref{fig:Fig4} shows the ensemble of all $KN\times1$ received
signals $\mathbf{y}\left(t\right)$ at time $t$ as given by

\textit{\tiny{}
\begin{equation}
\underset{\mathbf{y}\left(t\right)}{\underbrace{\left[\begin{array}{c}
\mathbf{y}_{1}\left(t\right)\\
\mathbf{y}_{2}\left(t\right)\\
\vdots\\
\mathbf{y}_{\mathrm{K}}\left(t\right)
\end{array}\right]}}=\underset{\mathbf{H}}{\underbrace{\left[\begin{array}{cccc}
\mathbf{H}_{11} & \mathbf{H}_{21} & \ldots & \mathbf{H}_{\mathrm{K}1}\\
\mathbf{H}_{12} & \mathbf{H}_{22} & \ldots & \mathbf{H}_{\mathrm{K}2}\\
\vdots & \vdots & \ddots & \vdots\\
\mathbf{H}_{1\mathrm{K}} & \mathbf{H}_{2\mathrm{K}} & \ldots & \mathbf{H}_{\mathrm{K}\mathrm{K}}
\end{array}\right]}}\underset{\mathbf{x}\left(t\right)}{\underbrace{\left[\begin{array}{c}
\mathbf{x}_{1}\left(t\right)\\
\mathbf{x}_{2}\left(t\right)\\
\vdots\\
\mathbf{x}_{\mathrm{K}}\left(t\right)
\end{array}\right]}}+\underset{\mathbf{w}\left(t\right)}{\underbrace{\left[\begin{array}{c}
\mathbf{w}_{1}\left(t\right)\\
\mathbf{w}_{2}\left(t\right)\\
\vdots\\
\mathbf{w}_{\mathrm{K}}\left(t\right)
\end{array}\right]}}\label{eq1}
\end{equation}
}where $\mathbf{x}_{k}\left(t\right)=\left[x_{k,1}\left(t\right),x_{k,2}\left(t\right)\ldots,x_{k,N}\left(t\right)\right]^{T}$
denotes the signals from the finite alphabet $\Lambda$ of M-QAM constellation
that is transmitted from the $N$ CPEs belonging to the $k$th SP.
The ensemble of $NK\times1$ transmitted signals related to all CPEs
from all the $K$ SPs are in $\mathbf{x}\left(t\right)$. Each $N\times N$
matrix $\mathbf{H}_{ij}$ is the $i\rightarrow j$ channel from $N$
CPEs of the $i$th SP ($\text{CP}\text{E}_{1,i},\ldots,\text{CP}\text{E}_{N,i}$)
towards the $j$th SP ($i\rightarrow j$) and it accounts for the
channel including the self-FEXT (when $i=j$) and alien-FEXT (when
$i\neq j$). $\mathbf{H}$ is the $KN\times KN$ channel matrix of
the multi-user/multi-operator system. The AWGN ${\bf w}_{k}(t)$ is
uncorrelated among CPEs and SPs, with the same power: $\mathbf{w}\left(t\right)\sim CN(0,\sigma^{2}\mathbf{I})$.
Since the proposed IC method handles mutual interference as useful
signals for the other SPs, for the scope of the paper the model (\ref{eq1})
needs to be re-ordered by grouping the CPEs that belong to the same
SP
\begin{equation}
\mathbf{y}\left(t\right)=\mathbf{H}_{1}\mathbf{x}_{1}\left(t\right)+\mathbf{H}_{2}\mathbf{x}_{2}\left(t\right)+\ldots+\mathbf{H}_{\mathrm{K}}\mathbf{x}_{\mathrm{K}}\left(t\right)+\mathbf{w}\left(t\right)\label{eq2}
\end{equation}
where ${\bf H}_{k}=\left[{\bf H}_{k1}^{T},{\bf H}_{k2}^{T},\ldots,{\bf H}_{kK}^{T}\right]^{T}$
is the compound channel from the $N$ CPEs belonging to the $k$th
SP ($\text{CP}\text{E}_{1,k},\ldots,\text{CP}\text{E}_{N,k}$) toward
\textit{all} the SPs as sketched in Fig.\ref{fig:Fig4}. The arrangement
(\ref{eq2}) will be used by the $k$th SP to estimate the $KN\times N$
channel ${\bf H}_{k}$ in IC channel estimation (Section III), or
in detection assuming ${\bf H}_{k}$ is known (Section IV). Model
for interference co-operation among the $K$ SPs is shown in Fig.\ref{fig:Fig4},
where all SPs share alien-FEXT interference information by using backhaul
for interference exchange that let every SP forward to \textit{all}
the $K-1$ SPs the received alien-interference after stripping the
own data. For $K=2$ it reduces to a single-link between SP1 and SP2.

To separate channel estimation from decoding beyond the logical separation,
the transmission is organized into frames with a frame structure that
alternates a set of training samples with data to be decoded. The
$k$th SP assigns to its $N$ CPEs $T$ samples of the training sequences
$\left\{ \overline{\mathbf{x}}_{k}\left(t\right)\right\} _{t=1}^{T}$
selected independently by the $k$th SP. Training used by the $k$th
SP are not known to the other SPs to prevent estimation of alien-FEXT
by other SPs. To ease the attention on IC method, training can be
considered as ideal (i.e., mutually orthogonal) even if some degradation
is expected when this condition is not met. Needless to say that the
channel estimation step includes the estimate of any frequency drift
and their correction for the own CPEs, if necessary (not covered here).
Channel estimation and MUD are discussed below as two distinct processing
steps, still sharing several commonalities due to the linear model
(\ref{eq2}).

\subsection{Self and Alien channel models}

Channel model depends on the cable length and frequency range as increasing
the frequency strengthen the inter-cable coupling, and this makes
the FEXT comparable with losses of direct links (insertion loss).
A simple model adopted here for numerical validations is based on
the assumption that the direct links are normalized to have unit-amplitude,
and the degree of diagonal dominance for self-FEXT channel $\mathbf{H}_{ii}$
and alien-FEXT channel $\mathbf{H}_{ij}$ (from $\{\text{CP}\text{E}_{\ell,i}\text{\}}_{\ell=1}^{N}$
to $\mathit{j}$th SP) is accounted in terms of scaling term $\alpha$,
the entries are
\begin{eqnarray}
{\bf H}_{ii}[p,p]=e^{j\theta_{p}}\text{ with }\theta_{p}\sim{\mathcal{U}}(0,2\pi)\\
{\bf H}_{ij}[p,q]\sim\mathcal{C}\mathcal{N}(0,\alpha^{2}),\forall p\ne q
\end{eqnarray}
with the same coupling $\alpha$ for both self- and alien-FEXT.

Off-diagonal terms are as small as $\alpha<-10dB$ up to 10-30MHz
bandwidth \cite{cit36} \cite{cit12}, these values are compliant
with channel measurements of 24 pairs (0.5mm/ea) over 200m twisted
pair cable up to 80-100MHz bandwidth as for next generation DSL \cite{cit37}.
The FEXT coupling $\alpha$ increases vs frequency and for long-cables,
it is not so unusual to have a FEXT that is comparable to direct link
when above 50-100MHz as for G.fast settings \cite{cit5} adopted in
numerical validations (Section V).

\section{Cooperative Channel Estimation}

The training sequences are composed of a set of $T$ complex-valued
samples ($T>N$) transmitted by all the $KN$ CPEs at the same time.
The $N\times T$ training sequence matrix for the $N$ CPEs of $k$th
SP \{$\text{CP}\text{E}_{\ell,k}\text{\}}_{\ell=1}^{N}$ is
\begin{equation}
\overline{\mathbf{X}}_{k}=\left[\overline{\mathbf{x}}_{k}\left(1\right),\overline{\mathbf{x}}_{k}\left(2\right),\ldots,\overline{\mathbf{x}}_{k}\left(T\right)\right],
\end{equation}
over-bar denotes the frame of training sequences. These training are
pseudo-random sequences assigned by SP to every CPE and never exchanged
among the SPs so the $k$th SP knows $\overline{\mathbf{X}}_{k}$,
but not the others. To simplify, it is assumed that the training sequences
of one SP are mutually uncorrelated with training sequences of other
$K-1$ SPs and have ideal auto-correlation so that $\overline{\mathbf{X}}_{k}\overline{\mathbf{X}}_{k}^{H}=\mathbf{I}$
and $\overline{\mathbf{X}}_{k}\overline{\mathbf{X}}_{i}^{H}=\mathbf{0}$
for all $i\neq k$. Notice that for random selection of the training
sequence as in practice happens, these ideal properties hold only
approximately at price of a negligible loss of performance (Section
V). The discrete signal received at $k$th SP ${\bf Y}_{k}=[{\bf y}_{k}(1),{\bf y}_{k}(2),\ldots,{\bf y}_{k}(T)]$
is
\begin{equation}
\mathbf{Y}_{k}=\mathbf{H}_{kk}\overline{\mathbf{X}}_{k}+\sum\limits _{m\neq k}\mathbf{H}_{mk}\overline{\mathbf{X}}_{m}+\mathbf{W}_{k}\label{eq3}
\end{equation}
here separated into own data ($\mathbf{H}_{kk}\overline{\mathbf{X}}_{k}$)
and alien-FEXT ($\sum\limits _{m\neq k}\mathbf{H}_{mk}\overline{\mathbf{X}}_{m}$).
Below the conventional Data Cooperation (DC) is discussed first as
reference scenario, and then the method based on Interference Cooperation
(IC).

\subsection{Data Cooperation (DC)}

In DC the initial signaling setup lets all SPs be aware of the training
sequence of all the other SPs. In this case the channel estimation
can be carried out by each SP to attain the centralized approach where
virtually one SP collects all the received signals. More specifically,
when the $k$th SP is aware of all training $\overline{\mathbf{X}}=\left[\overline{\mathbf{X}}_{1}^{T},\overline{\mathbf{X}}_{2}^{T},\ldots,\overline{\mathbf{X}}_{K}^{T}\right]^{T}$,
the model (\ref{eq3}) becomes:
\begin{equation}
\mathbf{Y}_{k}=\mathbf{G}_{k}^{T}\overline{\mathbf{X}}+\mathbf{W}_{k}
\end{equation}
where ${\bf G}_{k}^{T}=\left[{\bf H}_{1k},\cdots,{\bf H}_{kk},\ldots,{\bf H}_{Kk}\right]$
collects the channels from all $\text{CP}\text{E}_{\ell,i}$ toward
\textit{$k$}th SP. Maximum likelihood estimation (MLE) of this compound
channel is
\begin{equation}
\widehat{\mathbf{G}}_{k}=\mathbf{Y}_{k}\overline{\mathbf{X}}^{H}(\overline{\mathbf{X}}\overline{\mathbf{X}}^{H})^{-1}\label{eq3a}
\end{equation}
According to the properties of training sequence, the covariance of
the channel estimate $\widehat{\mathbf{G}}_{k}$ is
\begin{equation}
\operatorname{cov}(\widehat{{\bf G}}_{k})\geq\frac{\sigma^{2}}{T}{\bf I},\label{eq3b}
\end{equation}
that is the Cramér Rao bound of the channel estimate. These estimates
are independent and error scales with the training sequence length
$T$ and AWGN power $\sigma^{2}$. Once again, it is crucial to remark
that DC mimics the centralized approach so that $k$th SP estimates
self-FEXT (${\mathbf{H}_{kk}}$) and all the alien-FEXT (${\mathbf{H}_{mk}}$
for $\forall m\neq k$) channels that are locally used by $k$th SP
in MUD. If channel is slowly varying as in wired system, this method
needs a minimal inter-SP signaling at setup to have a consensus among
the usage of training and the channel estimations can be carried out
independently by each SP according to (\ref{eq3a}).

\subsection{Interference Cooperation (IC)}

IC method is iterative and implies not to exchange the training among
the SPs (as considered a sensitive information) but rather the training-induced
FEXT. The model for channel estimation by the $k$th SP as the SP
of interest is by collecting the $T$ training samples into $\mathbf{Y}=\left[\mathbf{y}(1),\mathbf{y}(2),\ldots,\mathbf{y}(T)\right]$
according to (\ref{eq2}):
\begin{equation}
\mathbf{Y}=\mathbf{H}_{k}\overline{\mathbf{X}}_{k}+\sum\limits _{m\neq k}\mathbf{H}_{m}\overline{\mathbf{X}}_{m}+\mathbf{W\mathrm{,}}\label{eq4a}
\end{equation}
where contribution by the different CPEs are grouped according to
their SPs. Since each SP is only aware of its own training sequence,
after the $k$th SP receives the alien-FEXT signaling from all the
other $K-1$ SPs in the form of $\mathbf{Y}_{m}-\sum\limits _{l\neq k}\mathbf{H}_{lm}\overline{\mathbf{X}}_{l}$
(say from the \textit{$m$}th SP with$m\neq k$), it stacks all the
received alien-FEXT to locally reproduce the following model complementary
to (\ref{eq4a})
\begin{equation}
\mathbf{Y}-\sum\limits _{m\neq k}\mathbf{H}_{m}\overline{\mathbf{X}}_{m}=\mathbf{H}_{k}\overline{\mathbf{X}}_{k}+\mathbf{W\mathrm{,}}\label{eq4}
\end{equation}
after the signals from all interfering SPs are stripped out from each
SP and forwarded to the \textit{$k$}th one. The linear model (\ref{eq4})
is solved iteratively in the IC method as detailed below.

Let $\mathbf{H}_{kk}^{(1)}$ be the initial estimate at $k$th SP
by assuming that alien-FEXT is an augmented AWGN, the $k$th SP exchanges
$\mathbf{Y}_{k}-\mathbf{H}_{kk}^{(1)}\overline{\mathbf{X}}_{k}$ with
the other $K-1$ SPs in trade of $\mathbf{Y}_{m}-\mathbf{H}_{mm}^{(1)}\overline{\mathbf{X}}_{m}$
one by one from all the others SPs (here $m\neq k$ denotes any arbitrary
SP different from the $k$th one). At the $n$th iteration of the
iterative IC method, the $k$th SP estimates $\mathbf{H}_{km}^{(n)}$
and this is re-encoded as $\mathbf{H}_{km}^{\left(n\right)}\overline{\mathbf{X}}_{k}$
and forwarded to the $m$th SP in form of alien-FEXT, and similarly
each of the other $K-1$ SPs re-encodes $\mathbf{H}_{mk}^{\left(n\right)}\overline{\mathbf{X}}_{m}$
for the benefit of $k$th SP. The interference is thus exchanged among
all the $K$ SPs one by one as $N\times T$ matrices. $k$th SP uses
the received re-encoded interference for alien-FEXT cancellation and
iterative channel estimation based on the linear model (\ref{eq4})
adapted for the iterations. It must be noticed that at every iteration
the estimate by the $k$th SP is the channel $KN\times N$ matrix
$\mathbf{H}_{k}=\left[\mathbf{H}_{k1}^{T},\mathbf{H}_{k2}^{T},\ldots,\mathbf{H}_{kK}^{T}\right]^{T}$
that stacks not only self-FEXT ($\mathbf{H}_{kk}^{(n)}$), but also
alien-FEXT channels from own CPEs towards the other SPs ($\mathbf{H}_{km}^{(n)}$
for $m\neq k$). At every iteration, the cost function is
\begin{equation}
\mathbf{H}_{k}^{\left(n+1\right)}=\arg\mathop{\min}\limits _{{\mathbf{H}_{k}}}\left\Vert \mathbf{Z}_{k}^{(n)}-\mathbf{H}_{k}\overline{\mathbf{X}}_{k}\right\Vert ^{2}
\end{equation}
where $\mathbf{Z}_{k}^{(n)}=\mathbf{Y}-\sum\limits _{m\neq k}\mathbf{H}_{m}^{\left(n\right)}\overline{\mathbf{X}}_{m}$
is the effective variable that collects all interference $\mathbf{H}_{m}^{\left(n\right)}\overline{\mathbf{X}}_{m}$
exchanged by the $K-1$ SPs ($m\neq k$) locally used for refinement
of $\mathbf{H}_{k}$. The estimation of $\mathbf{H}_{k}$ becomes
\begin{equation}
\mathbf{H}_{k}^{\left(n+1\right)}=\mathbf{Z}_{k}^{(n)}\overline{\mathbf{X}}_{k}^{H}(\overline{\mathbf{X}}_{k}\overline{\mathbf{X}}_{k}^{H})^{-1}
\end{equation}
Once again, SPs cooperation refines self-FEXT and alien-FEXT by constraining
the cooperation among SPs to never exchange the training $\overline{\mathbf{X}}_{k}$
and $\overline{\mathbf{X}}_{m}$ among each other, at most mixed by
the (unknown) channel responses. The iterative channel estimation
algorithm is summarized in Algorithm-1 for the $k$th SP.

\begin{algorithm}[tbh]
\protect\caption{Upstream Multi-operator Interference Cooperation Channel Estimation}

\begin{itemize}
\item Initialize ${\bf {H}}_{kk}^{\left(1\right)}={{\bf {Y}}_{k}}\overline{{\bf {X}}}_{k}^{H}{(\overline{{\bf {X}}}_{k}\overline{{\bf {X}}}_{k}^{H})^{-1}}$
\item Receive alien-FEXT ${{\bf {Y}}_{m}}-{\bf {H}}_{mm}^{\left(1\right)}{\overline{{\bf {X}}}_{m}}$
from all $K-1$ SPs ($m\ne k$).
\item Interference exchange ($m\leftrightarrow k$): evaluate ${\bf {H}}_{km}^{\left({1}\right)}=({{\bf {Y}}_{m}}-{\bf {H}}_{mm}^{\left({1}\right)}{\overline{{\bf {X}}}_{m}})\overline{{\bf {X}}}_{k}^{H}{(\overline{{\bf {X}}}_{k}\overline{{\bf {X}}}_{k}^{H})^{-1}}$
and forward ${\bf {H}}_{km}^{\left({1}\right)}{\overline{{\bf {X}}}_{k}}$
to $m$th SP in exchange of ${\bf {H}}_{mk}^{\left({1}\right)}{\overline{{\bf {X}}}_{m}}$.
\item for $n=1:{N_{iteration}}$

\begin{itemize}
\item Estimate \textit{
\[
{\bf {H}}_{kk}^{\left({n+1}\right)}=\left({{{\bf {Y}}_{k}}-\sum\limits _{m\neq k}{{\bf {H}}_{mk}^{(n)}{{\overline{{\bf {X}}}}_{m}}}}\right)\overline{{\bf {X}}}_{k}^{H}{({\overline{{\bf {X}}}_{k}}\overline{{\bf {X}}}_{k}^{H})^{-1}}
\]
}
\item Receive alien-FEXT ${{\bf {Y}}_{m}}-{\bf {H}}_{mm}^{\left({n+1}\right)}{\overline{{\bf {X}}}_{m}}-\sum\limits _{p\ne m,k}{{\bf {H}}_{pm}^{\left(n\right)}{{\overline{{\bf {X}}}}_{p}}}$
from all $K-1$ SPs ($m\ne k$).
\item Interference exchange ($m\leftrightarrow k$): evaluate ${\bf {H}}_{km}^{\left({n+1}\right)}=({{\bf {Y}}_{m}}-{\bf {H}}_{mm}^{\left({n+1}\right)}{\overline{{\bf {X}}}_{m}}-\sum\limits _{p\ne m,k}{{\bf {H}}_{pm}^{\left(n\right)}{{\overline{{\bf {X}}}}_{p}}})\overline{{\bf {X}}}_{k}^{H}{({\overline{{\bf {X}}}_{k}}\overline{{\bf {X}}}_{k}^{H})^{-1}}$
and forward ${\bf {H}}_{km}^{\left({n+1}\right)}{\overline{{\bf {X}}}_{k}}$
to $m$th SP in exchange of ${\bf {H}}_{mk}^{\left({n+1}\right)}{\overline{{\bf {X}}}_{m}}$. \end{itemize}
\end{itemize}
\end{algorithm}

\subsubsection{Convergence of IC}

Iterative IC method for channel estimation is equivalent to iteratively
solve the linear system of equations for the (centralized) model $\mathbf{Y}=\mathbf{H}\overline{\mathbf{X}}+\mathbf{W}$
from the (\ref{eq1}). In other word, given the centralized model
$\mathbf{Y}=\mathbf{H}\overline{\mathbf{X}}+\mathbf{W}$, the training
can be partitioned into self and alien-FEXT training ($\overline{\mathbf{X}}=\overline{\mathbf{X}}_{s}+\overline{\mathbf{X}}_{a}$)
so that the IC can be rewritten as ${\bf Y}={\bf H}^{(n+1)}\overline{{\bf X}}_{s}+{\bf H}^{(n)}\overline{{\bf X}}_{a}+{\bf W}$,
this is the basis to prove that the iterations in IC is equivalent
to Jacobi iteration \cite{Golub-book} for channel estimation such
that the estimate is carried out independently by each SP without
exchanging $\overline{\mathbf{X}}_{a}$ but rather in form of mixed
values ${\bf H}^{(n)}\overline{{\bf X}}_{a}$. Some technicalities
are necessary for the equivalence. Let $\mathbf{M}=blockdiag\left[\mathcal{X}_{1},\mathcal{X}_{2},\ldots,\mathcal{X}_{K}\right]$
be the block diagonal of the training sequences known by each SP,
where $\mathcal{X}_{k}=\mathbf{I}_{K}\otimes\overline{\mathbf{X}}_{k}^{T}\otimes\mathbf{I}_{N}$
here reordered just to comply with matrix algebra computations, the
block off-diagonal term $\mathbf{N}=\mathbf{M}-\left(\mathbf{1}_{K}\otimes\left[\mathcal{X}_{1},\mathcal{X}_{2},\ldots,\mathcal{X}_{K}\right]\right)$
denotes the matrix with training for alien-FEXT, the IC method (Algorithm
1) reduces to the set of Jacobi iterations
\begin{equation}
\mathbf{M}\cdot vec\left(\mathbf{H}^{(n+1)}\right)=\mathbf{N\cdot}vec\left(\mathbf{H}^{(n)}\right)+\left(\mathbf{1}_{K}\otimes vec(\mathbf{Y})\right)
\end{equation}
distributed over the SPs as each SP iteration refines (locally) a
portion of the channel estimate $vec\left(\mathbf{H}^{(n)}\right)$
where the estimates by each SP are arranged into a vector by $vec(.)$
operator. Proof of convergence of Jacobi iterations is in Appendix-A,
but since Jacobi iteration converge to $\mathbf{H}^{(n)}=\mathbf{Y}\overline{\mathbf{X}}^{H}(\overline{\mathbf{X}}\overline{\mathbf{X}}^{H})^{-1}$
for $n$ large enough (in practice, after 3-5 iterations, see Section
V) the IC converge to the centralized MLE of the channel matrix without
any exchange of data/training except as mixed values. It is crucial
to remark again that in IC method the channels ${\bf H}_{k}=\left[{\bf H}_{k1}^{T},{\bf H}_{k2}^{T},\ldots,{\bf H}_{kK}^{T}\right]^{T}$
that cause alien-FEXT towards the other remaining $K-1$ SPs is the
one that is updated within the $k$th SP. At convergence, this information
is resident in $k$th SP in form of channel estimates, and this enables
the IC Multi-user Detection (IC-MUD) as detailed below.

\section{Cooperative Multi-user Detection}

Data detection in multi-operator environment is based on channel state
information available at each SP and the degree of cooperation among
SPs. To ease the analysis still consistent with the problem at hand,
the channels are assumed as random and mutually orthogonal i.e. $E\left[\mathbf{H}_{ii}^{H}\mathbf{H}_{ij}\right]=0,\forall i\neq j$.
The signals received at $k$th SP for data detection follows from
(\ref{eq1}) (time index $t$ is omitted to simplify the notation):
\begin{equation}
\mathbf{y}_{k}=\mathbf{H}_{kk}\mathbf{x}_{k}+\sum\limits _{m\neq k}\mathbf{H}_{mk}\mathbf{x}_{m}+\mathbf{w}_{k}
\end{equation}
where $\mathbf{H}_{kk}\mathbf{x}_{k}$, $\mathbf{H}_{mk}\mathbf{x}_{m}$
and $\mathbf{w}_{k}$ accounts self-FEXT, alien-FEXT and AWGN. The
iterative MUD for Data Cooperation (DC) and Interference Cooperation
(IC) are based on the EM method \cite{cit10,cit11,kocian2003based,wu2008iterative},
and these iterative methods are compared to the centralized MUD (i.e.,
all the SPs are decoded jointly) used as a reference scenario.

Once defined the linear model, MUD can be based either on zero-forcing
(ZF) or minimum mean square (MMSE) criteria depending on the diagonal
dominance of the channel model \cite{cit4}. Herein all MUD methods
are based on the matrix decision-feedback equalizer (DFE) from QR
decomposition of the corresponding channel (or equivalently, BLAST
system) either using ZF or MMSE criteria as widely investigated and
adopted in vectoring (see e.g., \cite{Fischer-BDFE} and \cite{cit4}).
To simplify the notation of MUD, the metrics are referred to ZF-MUD
and matrix DFE are left indicated as well established.

\subsection{Centralized Multi-user Detection}

In centralized MUD all SPs forward the received signals to a central
processing unit that is aware of all channels $\mathbf{H}$ (possibly
estimated) and jointly decodes all the data streams according to the
system model (\ref{eq1}). The ZF-MUD is
\begin{equation}
\widehat{\mathbf{x}}=\mathbf{H}^{-1}\mathbf{y},
\end{equation}
and the variance of the noise at decision variable is lower bounded
by $\sigma^{2}/(1+(NK-1)\alpha^{2})$, it decreases with the number
of cooperating SPs ($K$) and CPEs ($N$) due to the augmentation
of signal flows in vectoring. For larger bandwidth (say >50MHz) the
FEXT-coupling increases and the MMSE-MUD with matrix-DFE takes more
efficiently into account the higher FEXT in vectoring. Signaling constraints
for data-rate and privacy issues make the full cooperation with centralized
MUD unfeasible but still it is an excellent reference bound in performance
analysis (Section V).

\subsection{Data Cooperation (DC) MUD}

In DC iterative technique, it is assumed that the channel $\mathbf{G}_{k}^{T}=\left[\mathbf{H}_{1k},\ldots,\mathbf{H}_{kk},\ldots,\mathbf{H}_{Kk}\right]$
is known to $k$th SP, or estimated as a separated step discussed
in Section III-A. At every iteration (say $n$th), every SP forwards
its decoded data $\mathbf{x}_{m}^{(n)}$ to all the other $K-1$ SPs
that in turn cancel locally the alien-FEXT using all the received
data from the $K-1$ SPs after channel reshaping $\mathbf{H}_{mk}\mathbf{x}_{m}^{\left(n\right)}$,
this step employs the following model for every iteration
\begin{equation}
\mathbf{y}_{k}-\sum\limits _{m\neq k}\mathbf{H}_{mk}\mathbf{x}_{m}^{(n)}=\mathbf{H}_{kk}\mathbf{x}_{k}+\mathbf{w}_{k}
\end{equation}
The ZF-MUD simplifies into a set of soft-detection for each decision
variable as \cite{cit10}:
\begin{equation}
\mathbf{x}_{k}^{(n)}=g_{\Lambda}\left[\frac{\mathbf{H}_{kk}^{-1}}{\sigma_{n}^{2}}\left(\mathbf{y}_{k}-\sum\limits _{m\neq k}\mathbf{H}_{mk}\mathbf{x}_{m}^{(n)}\right)\right]
\end{equation}
where $g_{\Lambda}[x]=E_{x\in\Lambda}\{x|x+z\}$ is the symbol-based
MMSE estimator at the decision variable and $\sigma_{n}^{2}$ is the
corresponding (iteration varying) variance of the noise at the decision
variable. The soft-decision $g_{\Lambda}[.]$ is modulation-dependent
as described in Appendix C, and it reduces to $\tanh[.]$ for BPSK
and QPSK modulations \cite{cit10}. Smoothness depends on $1/\sigma_{n}^{2}$
and it can be ``hardened'' (by reducing $\sigma_{n}^{2}$) at last
iterations, if necessary, or by updating its value as in \cite{cit27}.
The same soft-decisions are employed within the loop of the matrix-DFE
to avoid error propagation within the same SP.

At 1st iteration, data from other SPs are not available at any SP,
therefore from the received signal $\mathbf{y}_{k}$ the noise at
decision variable is assumed Gaussian and alien-FEXT augments the
AWGN: $\mathbf{H}_{kk}^{-1}(\sum\limits _{m\neq k}\mathbf{H}_{mk}\mathbf{x}_{m}+\mathbf{w}_{k})\sim CN\left(0,\sigma_{1}^{2}\mathbf{I}\right)$
with $\sigma_{1}^{2}=\left(\left(K-1\right)N\alpha^{2}+\sigma^{2}\right)/(1+(N-1)\alpha^{2})$.
Once the decoded data is shared among the SPs, the variance at decision
variable progressively reduces down to the (almost) complete alien-FEXT
cancellation $\sigma_{n}^{2}=\sigma^{2}/(1+(N-1)\alpha^{2})$, for
$n$ large enough. However, it must be noticed that the noise variance
in DC is always higher than the centralized approach as in DC only
the $N$ self-FEXT links are used for data estimation rather than
in centralized vectoring approach where all $KN$ links contributes
in decoding. Compared to centralized vectoring, the degradation of
noise power at decision variable for DC is at least
\begin{equation}
\text{DC}_{loss}=1+\frac{N(K-1)\alpha^{2}}{1+(N-1)\alpha^{2}}\label{eq8}
\end{equation}
that for a very large number of CPEs it depends on $K$ as DC$_{loss}\simeq K$.

\subsection{Interference Cooperation (IC) MUD}

For IC method, the reference system is (\ref{eq2}) and the ensemble
of channels ${\bf H}_{k}=\left[{\bf H}_{k1}^{T},{\bf H}_{k2}^{T},\ldots,{\bf H}_{kK}^{T}\right]^{T}$
is known to $k$th SP as being estimated by the IC channel estimation
(Section III-B). At each iteration, all the $K$ SPs estimates the
symbols $\mathbf{x}_{1}^{(n)},\mathbf{x}_{2}^{(n)},...,\mathbf{x}_{K}^{(n)}$
and forward each-others the received signals after stripping the own
signals and leaving the alien-FEXT. Focusing to the $k$th SP, it
stacks all the measurements received by all the K-1 SP into the $KN\times1$
vector
\begin{equation}
\mathbf{z}_{k}^{(n)}=\mathbf{y}-\sum\limits _{m\neq k}\mathbf{H}_{m}\mathbf{x}_{m}^{\left(n\right)}\label{eq:6a}
\end{equation}
that is the results of stripping $\mathbf{x}_{m}^{(n)}$ with the
local $KN\times N$ mixing channel$\mathbf{H}_{m}$ known only to
the $m$th SP ($m\neq k$) as detailed in Section III. The model at
the $k$th SP reduces to
\begin{equation}
\mathbf{z}_{k}^{(n)}=\mathbf{H}_{k}\mathbf{x}_{k}+\mathbf{w},\label{eq6}
\end{equation}
that is made redundant by $KN$ lines as the result of the augmentation
from the alien-FEXT exchange.

Signaling exchange and MUD processing for (\ref{eq6}) is detailed
in Algorithm 2 for ZF criteria (MMSE differs only for the metric),
and it is illustrated by focusing to a pair of SPs, say \textit{$m$}th
and \textit{$k$}th. The $m$th SP estimate $\mathbf{x}_{m}^{(n)}$
is mixed with (locally available) channel $\mathbf{H}_{mk}$ to have
$\mathbf{H}_{mk}\mathbf{x}_{m}^{(n)}$ that is forwarded to the $k$th
SP, and symmetrically from \textit{$k$}th SP so that the interference
is exchanged one-by-one among all the $K$ SPs. Based on these exchanges,
\textit{$k$}th SP mitigates the alien-FEXT from the locally received
signal $\mathbf{y}_{k}$ as $\mathbf{y}_{k}-\sum\limits _{m\neq k}\mathbf{H}_{mk}\mathbf{x}_{m}^{(n)}$.
Similarly, the $m$th SP estimates the alien-FEXT contribution from
$k$th SP as $\mathbf{y}_{m}-\sum\limits _{l\neq k}\mathbf{H}_{lm}\mathbf{x}_{l}^{(n)}$
using own CPEs ($\mathbf{H}_{mm}\mathbf{x}_{m}^{(n)}$) and alien-FEXT
($\mathbf{H}_{lm}\mathbf{x}_{l}^{(n)}$ for $l\neq k,m$) received
from the other SPs. The estimated alien-FEXT is again exchanged among
all the $K-1$ SPs and the $k$th SP to locally build the new $KN\times1$
vector (\ref{eq:6a}) for MUD ${\bf x}_{k}^{\left(n+1\right)}$.

After mutual alien-FEXT signaling, the estimation of $\mathbf{x}_{k}^{\left(n+1\right)}$
from the linear model (\ref{eq6}) is carried as part of the EM iterations
\cite{cit10,cit11}. Namely, the cost function for ZF-MUD (similarly
for MMSE-MUD) is given by
\begin{equation}
\mathbf{x}_{k}^{\left(n+1\right)}=\arg\min\limits _{\mathbf{x}_{k}}\left\Vert \mathbf{z}_{k}^{(n)}-\mathbf{H}_{k}\mathbf{x}_{k}\right\Vert ^{2}
\end{equation}
and it is be solved by matrix-DFE from QR decomposition of the compound
$KN\times N$ mixing channel$\mathbf{H}_{k}$ (see Algorithm 2 for
details). Due to the iterative exchange of alien-FEXT in interference
cancellation, it is convenient to adopt the soft-estimates to avoid
error propagation and have a gain in SNR of approx. 5dB (Section V)
in line with ref.\cite{cit27}. The MUD of the symbols at $(n+1)$th
iteration becomes:
\begin{equation}
\mathbf{x}_{k}^{\left(n+1\right)}=g_{\Lambda}\left[\frac{1}{\sigma_{n}^{2}}(\mathbf{H}_{k}^{H}\mathbf{H}_{k})\mathbf{H}_{k}^{H}\mathbf{z}_{k}^{(n)}\right]\label{eq7}
\end{equation}
where $\sigma_{n}^{2}$ is the variance of the noise at decision variable
for IC method and $g_{\Lambda}[.]$ is symbol-based MMSE estimator
based on the alphabet $\Lambda$ (Appendix C). Variance $\sigma_{n}^{2}$
reduces vs iterations and it can be calculated while evaluating the
new soft-decisions \cite{cit27}.

At the setup, there is no information available on alien-FEXT and
thus the estimate (\ref{eq7}) is replaced by the estimate based on
$\mathbf{y}_{k}$ that accounts for alien-FEXT as an augmented AWGN.
After the 1st iteration, IC and DC differ as received signals are
stripped of the own data to form the exchange signal $\mathbf{y}_{k}-\sum\limits _{m\neq k}\mathbf{H}_{mk}\mathbf{x}_{m}^{(n)}$
as detailed above (or in Algorithm 2). After alien-FEXT exchange,
the noise power at decision variable is bounded by $\sigma^{2}/(1+(NK-1)\alpha^{2})$
as for centralized approach.

{
\begin{algorithm}[tbh]
\protect\caption{Upstream Multi-operator Cooperative ZF-MUD}

\begin{itemize}
\item QR decomposition of $\mathbf{H}_{kk}$ (iteration 0): $\mathbf{Q\mathbf{R}}=\mathbf{H}_{kk}$,
and ${\bf \widetilde{y}}_{k}=\mathbf{Q}^{H}{\bf y}_{k}$.
\item Initialize
\[
\left[\mathbf{x}_{k}^{(1)}\right]_{i}=g_{\Lambda}\left[\frac{1}{\sigma_{n}^{2}}\left(\frac{1}{r_{i,i}}\left[{\bf \widetilde{y}}_{k}\right]_{i}-\sum\limits _{j=i+1}^{N}\frac{r_{i,j}}{r_{i,i}}\left[\mathbf{x}_{k}^{(1)}\right]_{j}\right)\right]
\]
, for $i=N,N-1,\ldots,1$
\item QR decomposition of $\mathbf{H}_{k}$ (iteration $n\geq1$):$\mathbf{Q\mathbf{R}}=\mathbf{H}_{k}$,
\item for $n=1:N_{iteration}$

\begin{itemize}
\item Receive ${\bf H}_{mk}{\bf x}_{m}^{\left(n\right)}$ from all$K-1$
SPs ($m\ne k$).
\item Interference exchange ($k\leftrightarrow m$): evaluate ${\bf y}_{k}-\sum\limits _{l\neq m}{\bf H}_{lk}{\bf x}_{l}^{\left(n\right)}$
and forward to $m$th SP in exchange of ${\bf y}_{m}-\sum\limits _{l\neq k}{\bf H}_{lm}{\bf x}_{l}^{\left(n\right)}$.
\item Estimate
\[
\left[\mathbf{x}_{k}^{(n+1)}\right]_{i}=
\]
\[
g_{\Lambda}\left[\frac{1}{\sigma_{n}^{2}}\left(\frac{1}{r_{i,i}}\left[\tilde{\mathbf{z}}_{k}^{(n)}\right]_{i}-\sum\limits _{j=i+1}^{N}\frac{r_{i,j}}{r_{i,i}}\left[\mathbf{x}_{k}^{(n+1)}\right]_{j}\right)\right]
\]
, for $\tilde{\mathbf{z}}_{k}^{(n)}=\mathbf{Q}^{H}\mathbf{z}_{k}^{(n)}$
\item Update $\sigma_{n}^{2}$ based on ${\bf x}_{k}^{\left(n+1\right)}$\end{itemize}
\end{itemize}
\end{algorithm}
}

\begin{figure}[h]
\centering \includegraphics[width=0.5\textwidth]{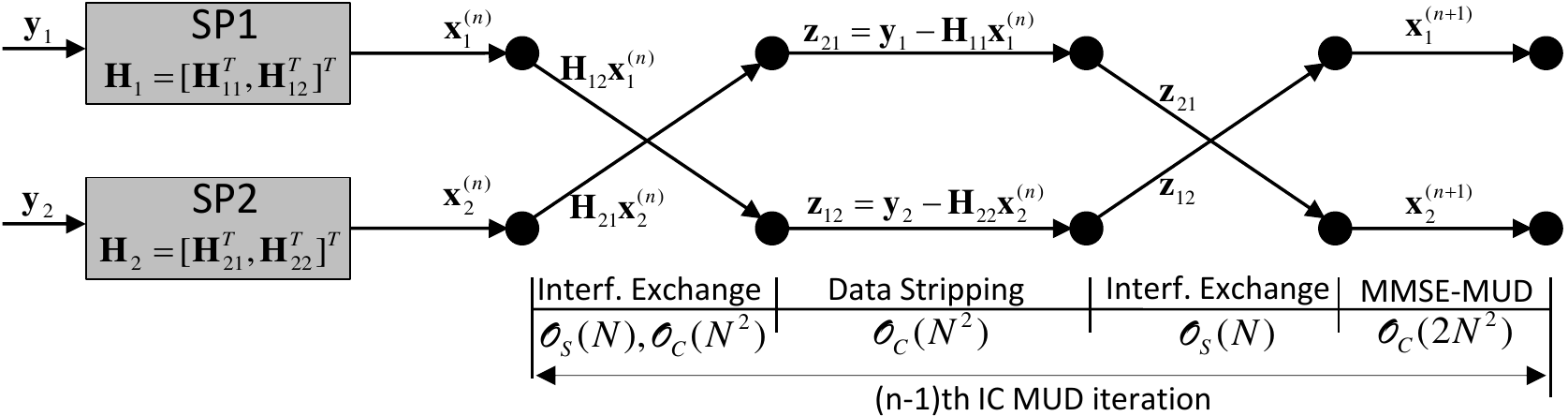}
\caption{Signaling and Computational Complexity at each SP for $K=2$.}

\label{fig:Fig5-1}
\end{figure}

Fig.\ref{fig:Fig5-1} illustrates the $(n+1)$th iteration of IC MUD
Algorithm 2 for $K=2$, there are two signaling phases with an overall
signaling complexity $\mathcal{O}_{S}\left(2N\right),$ and computational
complexity $\mathcal{O}_{C}\left(4N^{2}\right)$ at each SP. Generalization
to arbitrary $K$ SPs makes the overall complexity $\mathcal{O}_{S}\left(2\left(K-1\right)N\right)$
and $\mathcal{O}_{C}\left(2KN^{2}\right)$. Compared to DC MUD with
signaling complexity $\mathcal{O}_{S}\left(\left(K-1\right)N\right)$,
the signaling of IC MUD is twice with remarkable benefits (Section
V). Even if the data-rate is quite intensive, the SPs are hosted in
the same cabinet on at short distance, and there are several parallelism
that can be exploited in this setting for any practical implementations
\cite{cit39}.

\subsubsection{Convergence of ZF-MUD IC}

Similarly to channel estimation (as IC MUD rely on the same linear
model (\ref{eq1}) used in multi-user channel estimation), the updating
(\ref{eq7}) is conceptually equivalent to the partition of the channel
model into self ($\mathbf{H}_{s}$) and alien-FEXT ($\mathbf{H}_{a}$)
so that the iterations reflect this partitioning eased by the IC signaling,
and the local availability of channel matrices $\mathbf{H}_{k}$:
$\mathbf{y}=\mathbf{H}_{s}\mathbf{x}^{(n+1)}+\mathbf{H}_{a}\mathbf{x}^{(n)}+\mathbf{w.}$
The proof of convergence toward the centralized estimate follows the
same steps that highlights the equivalence between iterations (\ref{eq7})
and the Jacobi method to solve the linear system $\mathbf{y}=\mathbf{Hx{\textstyle +}w}$
\cite{Golub-book}. By reshaping the linear problem into block diagonal
matrix $\mathbf{M}=blockdiag\left[\mathbf{H}_{1},\mathbf{H}_{2},\ldots,\mathbf{H}_{K}\right]$
that accounts for self-FEXT (say $\mathbf{H}_{s}$), and block off-diagonal
matrix $\mathbf{N}=\mathbf{M}-\left(\mathbf{1}_{K}\otimes\left[\mathbf{H}_{1},\mathbf{H}_{2},\ldots,\mathbf{H}_{K}\right]\right)$
for alien-FEXT (say $\mathbf{H}_{a}$), each IC MUD is equivalent
to the Jacobi iteration:
\begin{equation}
\mathbf{M}\mathbf{x}^{(n+1)}=\mathbf{N}\mathbf{x}^{(n)}+\left(\mathbf{1}_{K}\otimes\mathbf{y}\right)
\end{equation}
carried out separately by each SP as MUD step (\ref{eq7}), except
for soft-detector $g_{\Lambda}[.]$ that is unavoidably related to
the alphabet $\Lambda$ and used to prevent error propagation. Once
again, the proof of convergence of Jacobi iterations (Appendix B)
guarantees that IC converges to the same centralized vectoring solution
for ZF metric. It can be shown that proof for MMSE-MUD IC follows
the same steps.

\subsubsection{Privacy}

The exchange between any two SP is mixed by alien-FEXT channel that
is never know according to the IC method for channel estimation (Section
III). The $k$th SP could recover the mixed signal $\mathbf{H}_{mk}\mathbf{x}_{m}^{(n)}$
but this is without the knowledge of the mixing alien-FEXT channel
$\mathbf{H}_{mk}$ that in IC method is not known to the \textit{$k$}th
SP, but only to the $m$th SP, and this randomizes $\mathbf{x}_{m}^{(n)}$
to preserve its the privacy \cite{Privacy Preserving}. Even if the
exchange among SPs is interference-based, in principle there could
be a way for the other SPs to infer the information on the own users.
For the example at hand, the estimation of the alien information $\mathbf{x}_{m}^{(n)}$
could reduce to a multi-user blind separation from $\mathbf{H}_{mk}\mathbf{x}_{m}^{(n)}$
with the unknown mixing matrix $\mathbf{H}_{mk}$. However, even if
the modulating terms are non-Gaussian as a necessary condition for
blind-estimation methods, these are unknown in term of constellation,
power (or scaling factor) and phase-stationarity (e.g., each user
can have an arbitrarily rotated constellation specifically employed
to prevent the decoding by alien SPs, but still being determistically
known for the own SP) to prevent a reliable decoding of $N\times N$
MIMO mixing that, for the alien-FEXT channel $\mathbf{H}_{mk}$, it
is likely below the decoding capability. In this sense the privacy
can be considered pragmatically preserved to enable the adoption of
IC methods from commercial operators and enable local loop unbundling.

\section{Numerical results}
\begin{figure*}[t!]
\centering \includegraphics[width=0.8\textwidth]{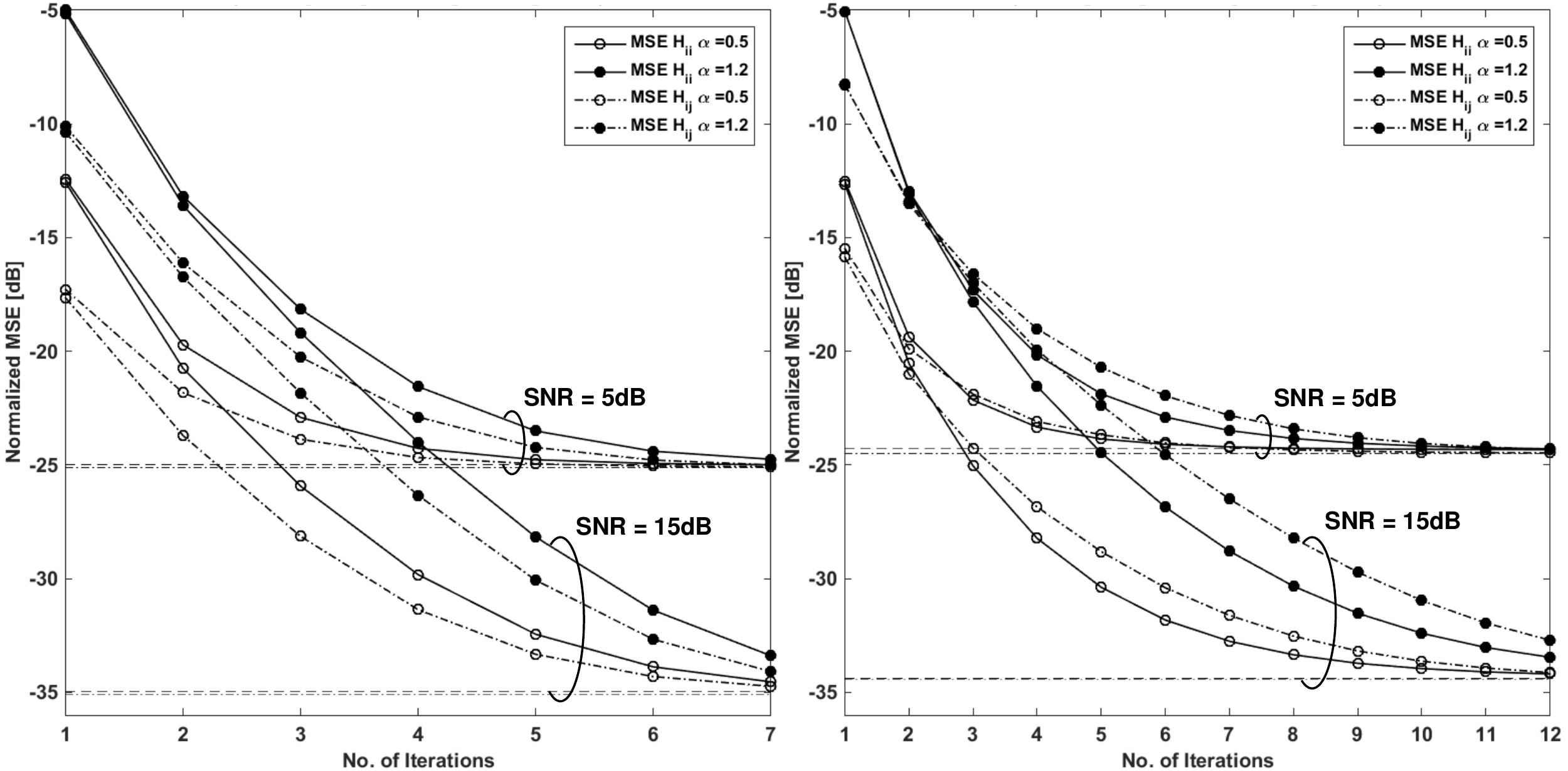}
\caption{MSE vs Iterations($n$) of IC channel estimation in multi-operator
xDSL for $SNR=[5,15]$dB,$\alpha=[0.5,1.2],K=2$ (left) and $K=3$
(right).}
\label{fig:Fig5}
\end{figure*}

Numerical simulations validate the asymptotic performance of the IC
method and evaluate the effectiveness of the cooperative method based
on the exchange of interference only. We consider a scenario with
$K=2$ and $3$ operators with $N=10$ CPEs each as being a reference
for G.fast setting. Performances of the IC algorithms are evaluated
either for statistical model \cite{G.fast-experimental} by varying
the FEXT coupling $\alpha$, and by considering the cable-models of
100m length \cite{cit37} up to the frequency of 200MHz to simulate
the behavior in G.fast settings. In any case, the direct link is independent
over the lines (all lines with the same length) so that $\mathbf{H}_{kk}[p,p]=\exp(j\theta_{p})$
with $\theta_{p}\sim\mathcal{U}(0,2\pi)$ statistically independent
over lines. The transmitted symbols from every CPE belong to QAM constellation
with a size that depends on the degree of G.fast specifications. Power
spectral density of the signal at CPE is -76dBm/Hz and the noise is
-140dBm/Hz as customary \cite{cit37,ITU-Simulation Condition,Choffi Book},
signal to noise ratio (SNR) is always referred at the receiver unless
defined at decision variable.

\subsection{Iterative Channel Estimation}

Training are generated from a random set of QPSK symbols known to
each SP as this approximates the ideal training with $\overline{\mathbf{X}}_{k}\overline{\mathbf{X}}_{k}^{H}\simeq\mathbf{I}$
and $\overline{\mathbf{X}}_{k}\overline{\mathbf{X}}_{i}^{H}\simeq\mathbf{0}$
for all $i\neq k$. Fig.\ref{fig:Fig5} shows MSE vs iterations of
self ($\mathbf{H}_{ii}$) and alien-FEXT ($\mathbf{H}_{ij}$) channel
estimates for $K=2$ and 3 operators scenario using IC, and the corresponding
Cramér Rao Bound (CRB) (\ref{eq3b}). The performance has been evaluated
for SNR$=1/\sigma^{2}\in\{5dB,15dB\}$ and FEXT coupling $\alpha\in\{0.5,1.2\}$
to validate the convergence of the proposed algorithms in these severe
FEXT-conditions. The MSE is normalized by the power of channel for
self-FEXT $\text{E}\left[\left\Vert \mathbf{H}_{ii}\right\Vert ^{2}\right]$
and alien-FEXT $\text{E}\left[\left\Vert \mathbf{H}_{ij}\right\Vert ^{2}\right]$;
since $\text{E}\left[\left\Vert \mathbf{H}_{ii}\right\Vert ^{2}\right]>\text{E}\left[\left\Vert \mathbf{H}_{ij}\right\Vert ^{2}\right]$
the normalized MSE for $\mathbf{H}_{ii}$ is lower than MSE of $\mathbf{H}_{ij}$.
Centralized method attains the CRB $MSE_{{\text{CRB}}}=trace\left[\sigma^{2}(\overline{\mathbf{X}}\overline{\mathbf{X}}^{H})^{-1}\right]$,
and here IC attains the CRB in 3-5 iterations for $K=2$. When $K>2$
a larger number of iterations is necessary as the iterations are implicitly
pairing the alien-FEXT to each users during the alien-FEXT exchange
steps. If FEXT-coupling $\alpha$ is smaller, the number of iterations
for IC convergence decreases as expected for smaller interference
settings. In G.fast scenario the IC method converges (with in +2-3dB
of excess MSE compared to CRB) in $2\div3$ iterations up to a frequency
of 100-120MHz, even if a practical convergence (not necessarily to
the CRB) within 1-2 iterations. The cost of inter-SP signaling in
this case is limited to the set-up phase and it needs not further
optimizations to comply with inter-SP data-rate.

\subsection{Iterative Detection}

\begin{figure}{h}
\centering \includegraphics[width=0.5\textwidth]{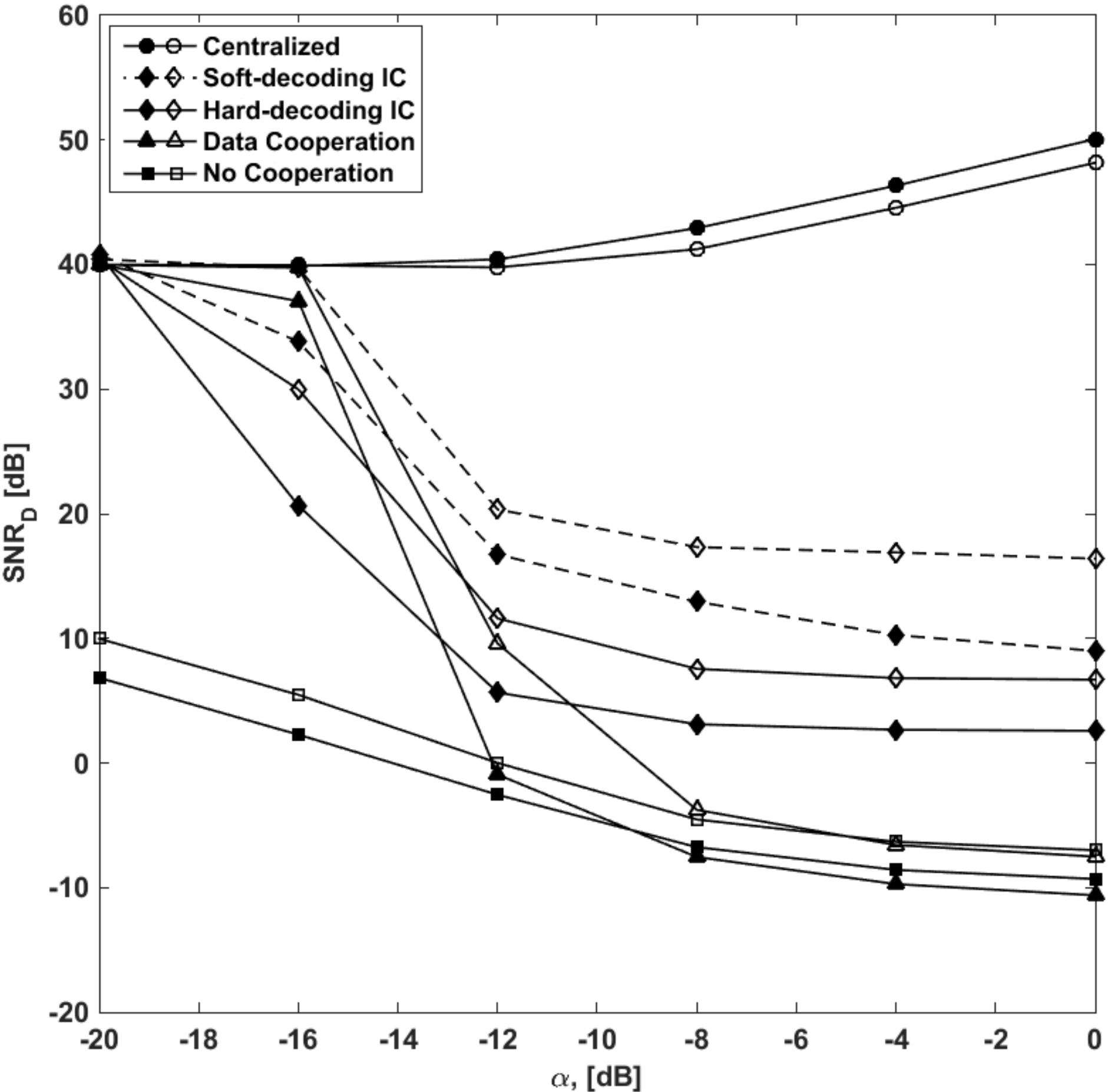}
\caption{SNR at decision variable of MMSE-MUD methods vs $\alpha$ for $K=2$
(empty marker) and $K=3$ (filled marker) with $SNR=40dB$ and 6 iterations.}

\label{fig:Fig6}
\end{figure}

Performance for multi-user/multi-operator detection are evaluated
in term of SNR at decision variable ($SNR_{D}$) for different FEXT
coupling and modulations. MMSE criteria for MUD is employed to cope
with large FEXT coupling. Fig.\ref{fig:Fig6} shows the $SNR_{D}$
for varying $\alpha$ and SPs $K={2,3}$ for SNR $=1/\sigma^{2}=15dB$
scenario after 6-iterations for DC and IC methods with QPSK modulation.
Performance of IC MMSE-MUD is compared with DC MMSE-MUD and MMSE-MUD
with local vectoring (without alien-FEXT compensation) thus showing
that the IC MMSE-MUD outperforms all the methods. Degradation with
respect to centralized MUD is mainly due to error propagation for
large FEXT coupling, soft-decoding in IC MUD prevents the propagation
of errors and guarantees $SNR_{D}\geq10dB$ even for $\alpha=0dB$
and $K=2,3$. Any inter-SP cooperation becomes mandatory for large
degree of coupling, say $\alpha>-20$dB, with IC MUD uniformly when
the error propagation dominates. Since in G.fast the crosstalk can
be considered as $\alpha\geq-20dB$, the cooperation among SP should
be considered as mandatory to make an efficient usage in any condition
of coexistence of users within the same cable bundling.

\begin{figure}{h}
\centering \includegraphics[width=0.5\textwidth]{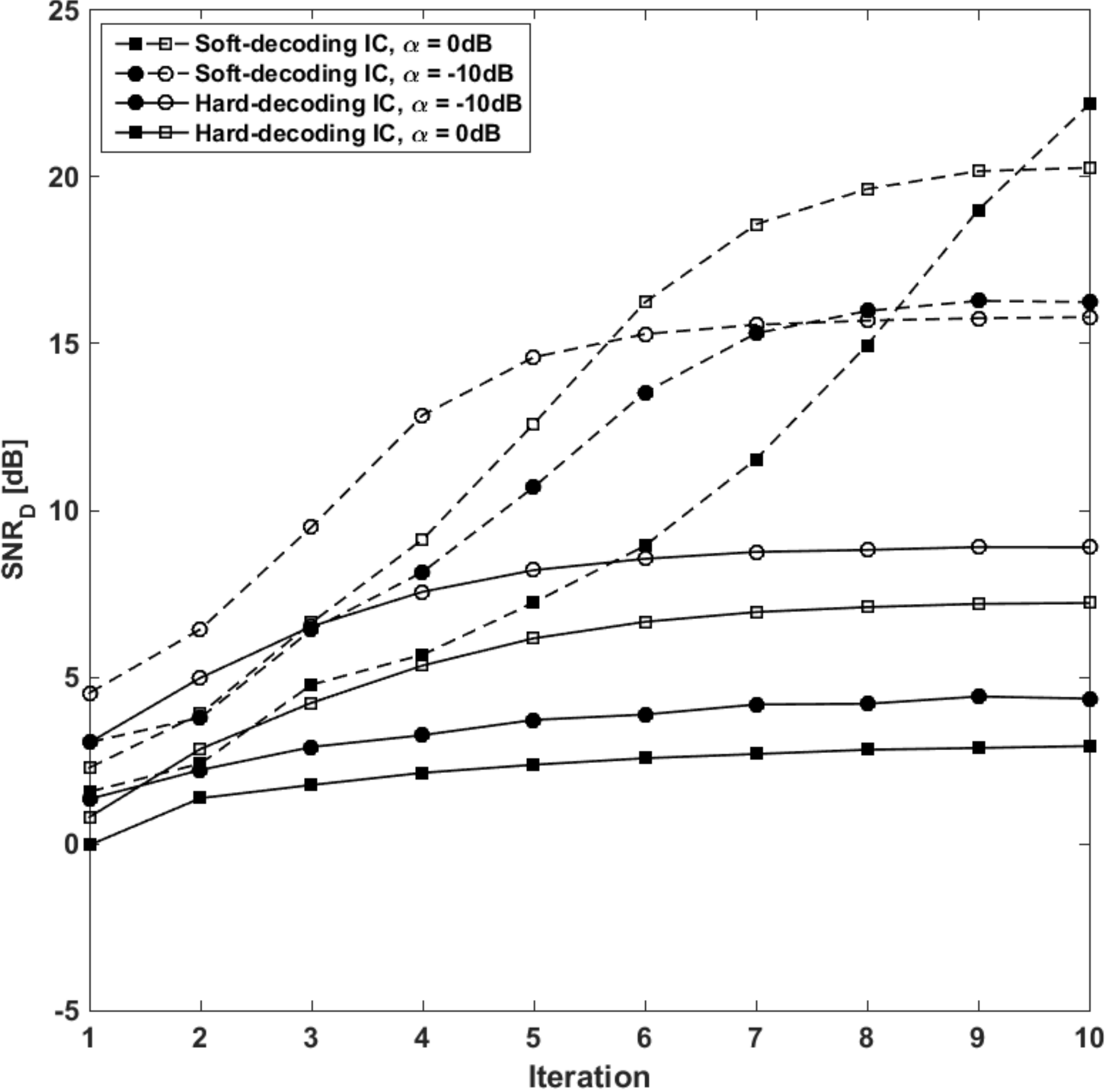}
\caption{SNR at decision variable vs iterations of IC MMSE-MUD for $K=2$ (empty
marker) and $K=3$ (filled marker), DFE with soft (dashed line) and
hard (solid line) decisions, and varying $\alpha=$\{-10(circle),0(square)\}
dB. $(N=10,SNR=15dB)$}

\label{fig:Fig7}
\end{figure}

Convergence analysis of IC MMSE-MUD for QPSK with soft (empty-dots)
and hard (filled dots) decisions DFE is in Fig.\ref{fig:Fig7} showing
the $SNR_{D}$ vs iterations for $K=2,3$ and varying $\alpha$. Once
again, this numerical results validates that the iterative IC algorithm
converges into few iterations (say 3-5 iterations) for most practical
purposes, with a remarkable benefit when using soft-decisions (\ref{eq7}).
Convergence is faster for smaller $\alpha$ and slower for large number
of SPs (here K=3) as iterations implicitly ease the assignment to
each SP the alien-FEXT corresponding to the own users to be exploited
efficiently as ``useful signal''.

\begin{figure}{h}
\centering \includegraphics[width=0.5\textwidth]{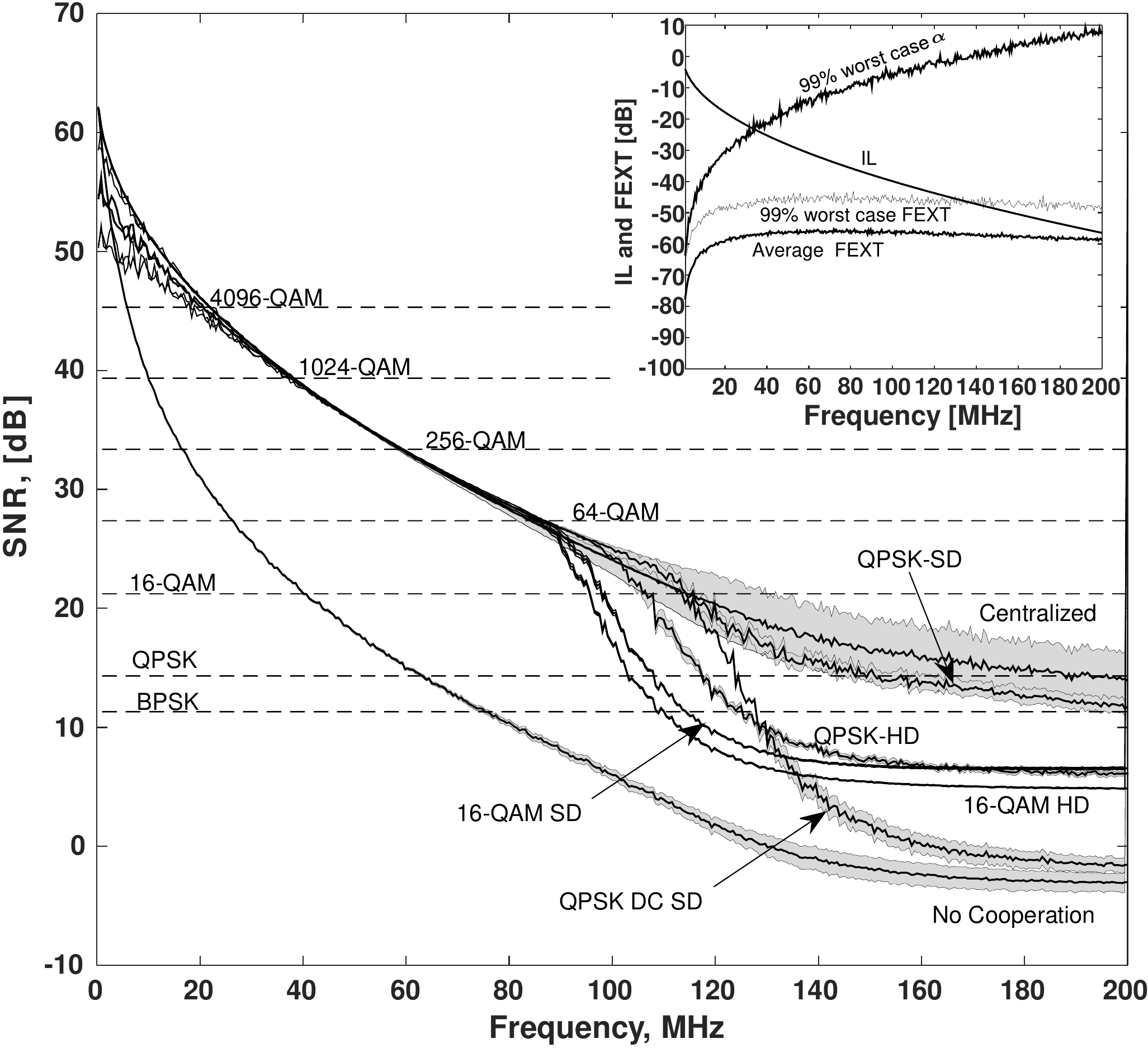}
\caption{SNR at decision variable vs frequency in G.fast settings (cable length
is 100m for all the CPEs, N=10 CPEs per SP and K=2 SPs) for MMSE-MUD:
IC method with soft (SD) and hard (HD) decoding, centralized MUD,
DC method. Channel losses are in top-right corner.}

\label{fig:Fig8}
\end{figure}

Average performance from all lines vs frequency for G.fast channel
model \cite{cit37,G.fast-experimental} (cable length is 100m for
all the CPEs, N=10 CPEs per SP and K=2 SPs) is in Fig.\ref{fig:Fig8},
channel model is sketched on top-corner for insertion-loss (IL) and
FEXT, and the corresponding $\alpha$. The thresholds on $SNR_{D}$
to attain the symbol error rate of $10^{-7}$ are superimposed for
different QAM constellations, from BPSK to 4096-QAM. The MMSE-MUD
for centralized system is the upper-bound with some dispersion of
values (shaded area) but still above the BPSK threshold. The lack
of any alien-FEXT mitigation as for no-cooperation scenario offers
the lower bound. The DC MMSE-MUD method shows comparable performance
with centralized MUD up to approx.120MHz (or equivalently from the
channel parameters $\alpha<-10dB$) in accordance with threshold effect
in Fig.\ref{fig:Fig6}. IC MMSE-MUD (4 iterations) with soft-decoding
outperforms the DC MUD and the same IC MUD with hard decoding for
QPSK and 16-QAM (other modulations are consistent with these results),
thus confirming all the results in Fig.\ref{fig:Fig6}. In addition,
the IC method guarantees the transport over all the G.fast bandwidth
up to 212MHz.

The SNR lets to compute the average throughput (Table 1) according
to the ``gap-formula'' for bit-loading usually adopted in DSL \cite{ITU-Simulation Condition,Choffi Book}
for $SNR_{D}$ at $\ell$th tone
\[
b_{\ell}=log_{2}\left(1+\frac{SNR_{D,\ell}}{\Gamma}\right)
\]
where the gap $\varGamma$=10.8dB=6dB(SNR margin)+9.8dB(SNR gap)-5dB(coding
gain) according to the ITU specifications for symbol error probability
$10^{-7}$, maximum loading of 12bits and framing overhead is 12\%
\cite{ITU-Simulation Condition}. Over the bandwidth 2-106MHz DC and
IC attain the same throughput of centralized MUD, but when extending
the bandwidth up to 212MHz the loss of IC MUD is negligible compared
to centralized method, and the DC looses approx. 100MHz in throughput.
The same Table shows that is it remarkably better to partition the
overall bandwidth among the 2 SPs (e.g., alternate usage of the tones
per SP) rather than allocate all the CPEs over the same bandwidth
without any degree of cooperation, in this case the throughput is
half the value for centralized (or IC) MUD.

\begin{table*}[tbh]
\protect\caption{Average throughput of different cooperation schemes for G. Fast settings
(N=10, K=2, cable:100m)}

\centering{}%
\begin{tabular}{|c|c|c|}
\hline
Cooperation Scheme & 2-106 MHz  & 2-212 MHz \tabularnewline
\hline
\hline
Centralized & 718 Mbps & 884 Mbps\tabularnewline
\hline
Equally shared bandwidth & 357 Mbps & 425 Mbps\tabularnewline
\hline
Interference Cooperation (IC) & 714 Mbps & 854 Mbps\tabularnewline
\hline
Data Cooperation (DC) & 711 Mbps & 756 Mbps\tabularnewline
\hline
No Cooperation (NC) & 275 Mbps & 275 Mbps\tabularnewline
\hline
\end{tabular}
\end{table*}

\section{Conclusions}

Conventional interference mitigation techniques are based on centralized
iterative interference cancellation that exchange data without any
limitation on ownership. In this paper, we propose multi-user receivers
employing channel estimation and detection in the multi-operator xDSL
unbundling scenario where cooperation among different operators is
obtained by exchanging the alien-interference considered as useful
signal by the other operators. The IC method is iterative as interference
exchange refines alien-interference and own data, and it has the merit
that this cooperation among different service providers never exchange
the decoded symbols considered as sensitive-information. Since each
operator receives the interference from the others in exchange of
their own, it can be considered an interference-based cooperation
that can exploit all the redundancy from all the data-paths with respect
to the other operators and thus the performance attains the same performance
as a centralized vectoring method.

The convergence of IC is impaired by error propagation and soft-decision
multi-user detection guarantees the convergence in few iterations
(type.3-5 iterations) with a threshold in loss when overall interference
becomes comparable with signal of interest. Even if data-rate exchange
among operators is double the interference mitigation method by data
exchange, the parallelism over frequency can be easily exploited in
inter-operator communication using specific architectures \cite{cit39}.

\appendices{}

\section{IC Channel Estimation Convergence}

The exchanging of the interference reduces the channel estimation
for $K$ SPs to the linear system (except AWGN that is irrelevant
for MLE method) $\mathbf{Y}=\mathbf{H}\overline{\mathbf{X}}$, after
vectorization this is rewritten as linear system
\begin{equation}
{\mathbf{b}}={\mathbf{Ah}}\label{eqA1}
\end{equation}
to be solved with respect to $\mathbf{h}=vec\left(\mathbf{H}\right)\in\mathbb{C}^{{K^{2}}{N^{2}}{\times1}}$.
Terms in (\ref{eqA1}) are ${\mathbf{b=1}_{K}}\otimes vec\left(\mathbf{Y}\right)\in\mathbb{C}^{{{K^{2}}NT}{\times1}}$,
${\mathbf{A=}{\mathbf{1}_{K}}\otimes\left[{{{\mathcal{X}}_{1}},{{\mathcal{X}}_{2}},\ldots,{{\mathcal{X}}_{K}}}\right]\in\mathbb{C}^{{K^{2}}NT\times{K^{2}}{N^{2}}}\mathbf{{\mathcal{\ }}}}$
where training for the \textit{k}th SP are reordered as ${\mathcal{X}_{k}}={\mathbf{I}_{K}}\otimes\overline{\mathbf{X}}_{k}^{T}\otimes{\mathbf{I}_{N}\in\mathbb{C}^{{K}NT\times{K^{2}}{N^{2}}}}$
and ${\mathbf{1}_{K}}={\left[{1,1,\ldots,1}\right]^{T}}$ of size
$K\times1$. Block-matrix $\mathbf{A}$ of training sequences can
be decomposed based on the knowledge of each SP as block-diagonal
$\mathbf{M}=blockdiag\left[{{{\mathcal{X}}_{1}},{{\mathcal{X}}_{2}},\ldots,{{\mathcal{X}}_{K}}}\right]$
(for each SP) and block off-diagonal matrix $\mathbf{N}=-(\mathbf{A}-\mathbf{M)}$
(signaled by the other SPs) so that the updates at every SP is equivalent
to the Jacobi iterations:
\begin{equation}
{\mathbf{M}}{{\mathbf{h}}^{(n+1)}}={\mathbf{N}}{{\mathbf{h}}^{(n)}}+{\mathbf{b}}
\end{equation}
to be applied in least-square sense as $\mathbf{A}$ is not a square
matrix:$\mathbf{h}$
\begin{equation}
{{\mathbf{h}}^{(n+1)}}={\left({{{\mathbf{M}}^{H}}{\mathbf{M}}}\right)^{-1}}{{\mathbf{M}}^{H}}{\mathbf{N}}{{\mathbf{h}}^{(n)}}+{\left({{{\mathbf{M}}^{H}}{\mathbf{M}}}\right)^{-1}\mathbf{b.}}
\end{equation}
Convergence to ${{\mathbf{h}}^{(\infty)}}={\left({{\mathbf{A}}^{H}\mathbf{A}}\right)^{-1}{\mathbf{A}}^{H}\mathbf{b}}$
regardless of the initialization ${{\mathbf{h}}^{(0)}}$ depends on
the spectral radius of ${\left({{\mathbf{M}^{H}}\mathbf{M}}\right)^{-1}}{\mathbf{M}^{H}}\mathbf{N}$
\cite{Golub-book}. Since $\mathbf{M}$ is diagonal dominant, the
spectral radius $\rho({\left({{\mathbf{M}^{H}}\mathbf{M}}\right)^{-1}}{\mathbf{M}^{H}}\mathbf{N})<1$,
and then the Jacobi iterations always converges for any starting vector
${\mathbf{h}^{(0)}}$. Proof is given below.

Let $\mathbf{e}{^{(n)}}={\mathbf{h}^{(n)}}-{\mathbf{h}^{(\infty)}}$
is the error at $n$th iteration. As $\mathbf{M}{\mathbf{h}^{(n+1)}}=\mathbf{N}{\mathbf{h}^{(n)}}+\mathbf{b}$,
the iterations are equivalent to the update $\mathbf{e}{^{(n+1)}}={\left({{{\mathbf{M}}^{H}}{\mathbf{M}}}\right)^{-1}}{{\mathbf{M}}^{H}}{\mathbf{Ne}^{(n)}}$
and convergence is guarantee if the eigenvalues of ${\left({{\mathbf{M}^{H}}\mathbf{M}}\right)^{-1}}{\mathbf{M}^{H}}\mathbf{N}$
are strictly smaller than 1. Rewriting the updating as
\begin{equation}
{\left({{{\mathbf{M}}^{H}}{\mathbf{M}}}\right)^{-1}}{{\mathbf{M}}^{H}}{\mathbf{N}}={\mathbf{I}}-{\left({{{\mathbf{M}}^{H}}{\mathbf{M}}}\right)^{-1}}{{\mathbf{M}}^{H}}{\mathbf{A}}
\end{equation}
where ${\mathbf{M}^{H}}\mathbf{M}=blockdiag\left[{{\mathcal{X}}_{1}^{H}{{\mathcal{X}}_{1}},{\mathcal{X}}_{2}^{H}{{\mathcal{X}}_{2}},\ldots,{\mathcal{X}}_{K}^{H}{{\mathcal{X}}_{K}}}\right]$,
\begin{equation}
{{\mathbf{M}}^{H}}{\mathbf{A}}=\left[{\ \begin{array}{cccc}
{{\mathcal{X}}_{1}^{H}{{\mathcal{X}}_{1}}} & {{\mathcal{X}}_{1}^{H}{{\mathcal{X}}_{2}}} & \cdots & {{\mathcal{X}}_{1}^{H}{{\mathcal{X}}_{K}}}\\
{{\mathcal{X}}_{2}^{H}{{\mathcal{X}}_{1}}} & {{\mathcal{X}}_{2}^{H}{{\mathcal{X}}_{2}}} & \cdots & {{\mathcal{X}}_{2}^{H}{{\mathcal{X}}_{K}}}\\
\vdots & \vdots & \ddots & \vdots\\
{{\mathcal{X}}_{K}^{H}{{\mathcal{X}}_{1}}} & {{\mathcal{X}}_{K}^{H}{{\mathcal{X}}_{2}}} & \cdots & {{\mathcal{X}}_{K}^{H}{{\mathcal{X}}_{K}}}
\end{array}}\right],
\end{equation}
the update ${\left({{\mathbf{M}^{H}}\mathbf{M}}\right)^{-1}}{\mathbf{M}^{H}}\mathbf{A}$
becomes: \textit{
\[
{\left({{{\mathbf{M}}^{H}}{\mathbf{M}}}\right)^{-1}}{{\mathbf{M}}^{H}}{\mathbf{A}}=
\]
}

\textit{\tiny{}
\begin{equation}
\left[{\begin{array}{cccc}
{\mathbf{I}} & {{{\left({{\mathcal{X}}_{1}^{H}{{\mathcal{X}}_{1}}}\right)}^{-1}}{\mathcal{X}}_{1}^{H}{{\mathcal{X}}_{2}}} & \cdots & {{{\left({{\mathcal{X}}_{1}^{H}{{\mathcal{X}}_{1}}}\right)}^{-1}}{\mathcal{X}}_{1}^{H}{{\mathcal{X}}_{K}}}\\
{{{\left({{\mathcal{X}}_{2}^{H}{{\mathcal{X}}_{2}}}\right)}^{-1}}{\mathcal{X}}_{2}^{H}{{\mathcal{X}}_{1}}} & {\mathbf{I}} & \cdots & {{{\left({{\mathcal{X}}_{2}^{H}{{\mathcal{X}}_{2}}}\right)}^{-1}}{\mathcal{X}}_{2}^{H}{{\mathcal{X}}_{K}}}\\
\vdots & \vdots & \ddots & \vdots\\
{{{\left({{\mathcal{X}}_{K}^{H}{{\mathcal{X}}_{K}}}\right)}^{-1}}{\mathcal{X}}_{K}^{H}{{\mathcal{X}}_{1}}} & {{{\left({{\mathcal{X}}_{K}^{H}{{\mathcal{X}}_{K}}}\right)}^{-1}}{\mathcal{X}}_{K}^{H}{{\mathcal{X}}_{2}}} & \cdots & {\mathbf{I}}
\end{array}}\right]
\end{equation}
}{\tiny \par}

Since the training sequences are orthogonal (i.e., ${\overline{\mathbf{X}}_{k}}\overline{\mathbf{X}}_{k}^{H}=\mathbf{I}$
and ${\overline{\mathbf{X}}_{k}}\overline{\mathbf{X}}_{i}^{H}=\mathbf{0}$
for all $i\neq k$), or at least uncorrelated, the entries of the
off-diagonal blocks ${\left({{\mathcal{X}}_{k}^{H}{{\mathcal{X}}_{k}}}\right)^{-1}}{\mathcal{X}}_{k}^{H}{{\mathcal{X}}_{m}}$
are very small (ideally null). Hence the matrix is diagonal dominant
and spectral radius is $\rho({\left({{\mathbf{M}^{H}}\mathbf{M}}\right)^{-1}}{\mathbf{M}^{H}}\mathbf{N})<1$,
the convergence is guaranteed.

\section{IC Data Detection Convergence Conditions}

The exchanging of the interference reduces the detection for K SPs
to the linear system ${\bf {y}}={\bf {Hx}}$, after vectorization
it is
\begin{equation}
{\bf {b}}={\bf {Ax}}\label{eqB1}
\end{equation}
to be solved with respect to ${\bf {x}}\in{{\Bbb C}^{KN\times1}}$.
Terms in (\ref{eqB1}) are ${\bf {b}}={{\bf {1}}_{K}}\otimes{\bf {y}}\in{{\Bbb C}^{{K^{2}}N\times1}}$,
${\bf {A}}={{\bf {1}}_{K}}\otimes{\bf {H}}\in{{\Bbb C}^{{K^{2}}N\times KN}}$,
and ${\mathbf{1}_{K}}={\left[{1,1,\ldots,1}\right]^{T}}$ of size
$K\times1$. Block-matrix A of channel can be decomposed based on
the knowledge of each SP as block-diagonal $\mathbf{M}=blockdiag\left[{{\mathbf{H}_{1}},{\mathbf{H}_{2}},\ldots,{\mathbf{H}_{K}}}\right]$
and block off-diagonal matrix $\mathbf{N}=\mathbf{M}-\mathbf{A}$
so that the updates at every SP is equivalent to the Jacobi iterations:
\begin{equation}
{\mathbf{M}}{{\mathbf{x}}^{(n+1)}}={\mathbf{N}}{{\mathbf{x}}^{(n)}}+{\mathbf{b}}
\end{equation}
Similarly to Appendix-A, the iterative method is
\begin{equation}
{{\mathbf{x}}^{(n+1)}}={\left({{{\mathbf{M}}^{H}}{\mathbf{M}}}\right)^{-1}}{{\mathbf{M}}^{H}}{\mathbf{N}}{{\mathbf{x}}^{(n)}}+{\left({{{\mathbf{M}}^{H}}{\mathbf{M}}}\right)^{-1}}{{\mathbf{M}}^{H}}{\mathbf{b}}
\end{equation}
System convergence to ${{\bf {x}}^{(\infty)}}={\left({{{\bf {A}}^{H}}{\bf {A}}}\right)^{-1}}{{\bf {A}}^{H}}{\bf {b}}$
that for the structure of the problem coincides with ZF MUD (18),
and depends upon the spectral radius of ${\left({{\mathbf{M}^{H}}\mathbf{M}}\right)^{-1}}{\mathbf{M}^{H}}\mathbf{N}$.
The structure of the matrix is:
\begin{equation}
{\left({{{\mathbf{M}}^{H}}{\mathbf{M}}}\right)^{-1}}{{\mathbf{M}}^{H}}{\mathbf{N}}={\mathbf{I}}-{\left({{{\mathbf{M}}^{H}}{\mathbf{M}}}\right)^{-1}}{{\mathbf{M}}^{H}}{\mathbf{A}}
\end{equation}
where ${{\mathbf{M}}^{H}}{\mathbf{M}}=blockdiag\left[{{\mathbf{H}}_{1}^{H}{{\mathbf{H}}_{1}},{\mathbf{H}}_{2}^{H}{{\mathbf{H}}_{2}},\ldots,{\mathbf{H}}_{K}^{H}{{\mathbf{H}}_{K}}}\right]$
and
\begin{equation}
{{\mathbf{M}}^{H}}{\mathbf{A}}=\left[{\begin{array}{cccc}
{{\mathbf{H}}_{1}^{H}{{\mathbf{H}}_{1}}} & {{\mathbf{H}}_{1}^{H}{{\mathbf{H}}_{2}}} & \cdots & {{\mathbf{H}}_{1}^{H}{{\mathbf{H}}_{K}}}\\
{{\mathbf{H}}_{2}^{H}{{\mathbf{H}}_{1}}} & {{\mathbf{H}}_{2}^{H}{{\mathbf{H}}_{2}}} & \cdots & {{\mathbf{H}}_{2}^{H}{{\mathbf{H}}_{K}}}\\
\vdots & \vdots & \ddots & \vdots\\
{{\mathbf{H}}_{K}^{H}{{\mathbf{H}}_{1}}} & {{\mathbf{H}}_{K}^{H}{{\mathbf{H}}_{2}}} & \cdots & {{\mathbf{H}}_{K}^{H}{{\mathbf{H}}_{K}}}
\end{array}}\right]
\end{equation}
thus ${\left({{\mathbf{M}^{H}}\mathbf{M}}\right)^{-1}}{\mathbf{M}^{H}}\mathbf{A}$
becomes: \textit{
\[
{\left({{{\mathbf{M}}^{H}}{\mathbf{M}}}\right)^{-1}}{{\mathbf{M}}^{H}}{\mathbf{A}}=
\]
}

\textit{\tiny{}
\begin{equation}
\left[{\begin{array}{cccc}
{\mathbf{I}} & {{{\left({{\mathbf{H}}_{1}^{H}{{\mathbf{H}}_{1}}}\right)}^{-1}}{\mathbf{H}}_{1}^{H}{{\mathbf{H}}_{2}}} & \cdots & {{{\left({{\mathbf{H}}_{1}^{H}{{\mathbf{H}}_{1}}}\right)}^{-1}}{\mathbf{H}}_{1}^{H}{{\mathbf{H}}_{K}}}\\
{{{\left({{\mathbf{H}}_{2}^{H}{{\mathbf{H}}_{2}}}\right)}^{-1}}{\mathbf{H}}_{2}^{H}{{\mathbf{H}}_{1}}} & {\mathbf{I}} & \cdots & {{{\left({{\mathbf{H}}_{2}^{H}{{\mathbf{H}}_{2}}}\right)}^{-1}}{\mathbf{H}}_{2}^{H}{{\mathbf{H}}_{K}}}\\
\vdots & \vdots & \ddots & \vdots\\
{{{\left({{\mathbf{H}}_{K}^{H}{{\mathbf{H}}_{K}}}\right)}^{-1}}{\mathbf{H}}_{K}^{H}{{\mathbf{H}}_{1}}} & {{{\left({{\mathbf{H}}_{K}^{H}{{\mathbf{H}}_{K}}}\right)}^{-1}}{\mathbf{H}}_{K}^{H}{{\mathbf{H}}_{2}}} & \cdots & {\mathbf{I}}
\end{array}}\right]
\end{equation}
}{\tiny \par}

Since the off-diagonal terms are random (3-5) and ${\bf {H}}_{k}^{H}{{\bf {H}}_{m}}=\sum\limits _{l=1}^{K}{{\bf {H}}_{kl}^{H}{{\bf {H}}_{ml}}}\simeq0$
for large $N$ and $K$, ${\left({\mathbf{H}_{k}^{H}{\mathbf{H}_{k}}}\right)^{-1}}\mathbf{H}_{k}^{H}{\mathbf{H}_{m}}$
is close to 0. Hence ${\left({{\mathbf{M}^{H}}\mathbf{M}}\right)^{-1}}{\mathbf{M}^{H}}\mathbf{N}$
is diagonal dominant with $\rho({\left({{\mathbf{M}^{H}}\mathbf{M}}\right)^{-1}}{\mathbf{M}^{H}}\mathbf{N})<1$
and convergence is guaranteed.

\section{MMSE Estimator $g_{\Lambda}${[}.{]}}

Soft-detector $g_{\Lambda}[.]$ plays a role to avoid error propagation
and it depends on the transmitted constellation $\Lambda$. During
iterations, each symbol for each user at decision variable can be
modeled as $y=x+z$: the sum of a complex valued symbol $x\in\Lambda$
and a Gaussian noise $z\sim CN(0,\sigma^{2})$ with power $\sigma^{2}$
that depends on the iteration. The conditional expectation $g[y]=E\{x|y\}$
depends on the probability density function (pdf) $p_{y}[y]$ of $y$,
and in turn on pdf of $x$ and $z$ as being both random variable
statistically independent
\begin{equation}
g_{\Lambda}[y]=y+\sigma^{2}\frac{d}{dy}\log\{p_{y}[y]\}.
\end{equation}
For separable rectangular constellation (e.g., M-QAM: $\Lambda={\Lambda_{R}}\times{\Lambda_{I}}$
with ${\Lambda_{R}}={\Lambda_{I}}=\{\pm1,\pm3,..,\pm(2\sqrt{M}-1)\}$)
the soft-detector is separable $g_{\Lambda}[y]=\phi\lbrack y_{R}]+j\phi\lbrack y_{I}]$
onto real and imaginary component, and it resembles a soft-detector
for multilevel constellations and it becomes hard-detector for $\sigma^{2}\rightarrow0$.

To simplify, let $x\in\{\alpha_{1},\alpha_{2},...,\alpha_{M}\}$ where
$M$ symbols of the alphabet are equally likely, and $G(\zeta,\sigma^{2})=(2\pi\sigma^{2})^{-1/2}\exp(-\zeta^{2}/2\sigma^{2})=p_{z}[\zeta]$
be the Gaussian pdf of noise, the conditional mean becomes
\begin{equation}
\phi\lbrack y]=y+\sigma^{2}\frac{d}{dy}\log\left\{ \frac{1}{K}\sum_{k=1}^{K}G(y-\alpha_{k},\sigma^{2})\right\}
\end{equation}
and it reduces to $\phi\lbrack y]=\tanh[y/\sigma^{2}]$ for $x\in\{-1,+1\}$
(BPSK constellation).
\end{document}